\documentclass{article}
\usepackage{amssymb}
\usepackage[utf8]{inputenc}
\usepackage{textcomp}
\usepackage{hyperref}
\usepackage{graphicx}
\usepackage{amsmath}
\usepackage[version=4]{mhchem}
\usepackage{siunitx}
\usepackage{longtable,tabularx}
\usepackage{comment}
\usepackage{graphicx}
\usepackage{epstopdf, epsfig}
\usepackage{xcolor}
\usepackage{ulem}
\usepackage{mathtools}
\usepackage[numbers]{natbib}
\usepackage{graphicx}
\usepackage{epstopdf, epsfig}
\usepackage{bm}
\usepackage{subcaption}
\usepackage{float}
\usepackage{multirow}
\usepackage{booktabs}

\usepackage{color}
\definecolor{nCol}{rgb}{0.2, 0.7, 0}
\definecolor{lCol}{rgb}{0.0, 0.0, 0}
\usepackage[utf8]{inputenc}

\title{Deep learning-based predictive modelling of transonic flow over an aerofoil}

\author{Li-Wei Chen, Nils Thuerey \\ Technical University of Munich, D-85748 Garching
\\ jilinchl@163.com}
\date{22 March 2024}

\begin{document}

\maketitle

\begin{abstract}
%Efficiently predicting transonic unsteady flow over an aerofoil is an inherently challenging endeavor. In this study, we leverage deep neural network (DNN) models based on the attention U-Net architecture, 
%training them at efficient spatial and temporal resolutions 
Effectively predicting transonic unsteady flow over an aerofoil poses inherent challenges. In this study, we harness the power of deep neural network (DNN) models using the attention U-Net architecture. 
Through efficient training of these models,
we achieve the capability to capture the complexities of transonic and unsteady flow dynamics at high resolution, even when faced with previously unseen conditions.
We demonstrate that by leveraging the differentiability inherent in neural network representations, our approach provides a framework for assessing fundamental physical properties via global instability analysis.
This integration bridges deep neural network models and traditional modal analysis, offering valuable insights into transonic flow dynamics and enhancing the interpretability of neural network models in flowfield diagnostics.

\end{abstract}

\section{Introduction}
Self-sustained shock wave motions on aerofoils are closely linked with the occurrence of buffeting, causing pronounced shock-induced fluctuations. Several comprehensive reviews have shed light on this topic, including works by \cite{Nieuwland_Spee1973transonic}, \cite{Tijdeman_Seebass1980transonic} and \cite{Lee2001self}.
%%%%% above: just general overview %%%%%
%%%%% Buffet - Challenging %%%%%
%%%%% Global Instability
%%%%% Resolvent analysis %%%
%%%%% Learned coarse model %%%% 
To elucidate the underlying mechanism of self-sustained shock wave motion, \cite{Lee1990feedbackModel} proposed a phenomenological model that considers a closed feedback loop between the mean position of the shock wave and the trailing edge, which serves as the acoustic source. This feedback model has gained widespread recognition and has been extensively employed to scrutinize transonic flows over aerofoils \cite[]{Oliver2003Airfoil, Xiao2004buffet, Deck2005buffet, Chen2010DES}.
%%%%%%%%%%%%%%%%%%%%%%
% 
%In their study,

More quantitatively, \cite{crouch_garbaruk_magidov_travin2009origin} highlighted the significance of global instability in the onset of the transonic buffet. Building upon this, 
\cite{Sartor2015stability} further explored the impact of various disturbances and perturbations on buffet mode stability. Their investigation uncovered medium-frequency unsteadiness resembling Kelvin-Helmholtz-type instability, in addition to the well-documented low-frequency shock unsteadiness. 
Furthermore, \cite{kojima_yeh_taira_kameda_2020} utilized resolvent analysis to connect the buffet mechanism to the shock foot, shedding light on low-Reynolds number compressible aerodynamics and the influence of perturbations in the vicinity of the trailing edge.
The existence of both low and high frequencies in the context has been supported by the research conducted by 
\cite{Bouhadji2003a} and \cite{Bouhadji2003b}. They investigated unsteady transonic flows around a two-dimensional aerofoil, revealing four transition stages, including steady flow, spontaneous unsteadiness, a critical Mach number range, and two instability processes, with a focus on the influence of compressibility effects on these transitions. 
To explain the coexistence of the unsteadiness, they utilized Landau's model as an interpretive framework.

In addition to the aforementioned advancement of theoretical models, data-driven modeling methods demonstrate significant potential for modeling and analyzing transonic flows over aerofoils, encompassing techniques such as proper orthogonal decomposition \citep{Chen2010DES, moise_zauner_sandham_2022}, dynamic mode decomposition \citep{Kou2016, Weiner_Semaan2023robustDMD}, and data-driven resolvent analysis \citep{Iwatani2022reviewPODDMD, Iwatani_Asada_Yeh_Taira2023resolvent}. Those modal approaches facilitate the discernment of physically critical features within the complexity of high-dimensional and nonlinear phenomena. However, their limitations are also well-documented, as they can struggle with accurately predicting the behavior of highly nonlinear systems \citep{Taira2017reviewModalAnalysis}.
% our model is not interpretable.
% (cite 5 papers on dmd pod)
%%%%%%% Neural network and data-driven

Recently, the application of deep learning models for predicting the unsteady evolution of complex fluid dynamics has garnered significant attention and opened up new possibilities for performing data-driven predictive modeling of fluid dynamics. Some researchers have harnessed the potential of deep learning in conjunction with numerical solvers to effectively predict the non-stationary characteristics of turbulent flows \citep{Maulik2019, um2020sol, MacArt2021, Kochkov2021pnas, list_chen_thuerey_2022}. 
In contrast, others have pursued end-to-end supervised training and achieved remarkable outcomes
\citep{wang2020physicsinformed, stachenfeld2022learnedcoarse, allen2022inverse}. 
Nonetheless, a distinct gap persists in the availability of neural network-based predictive models for unsteady transonic flows, especially in situations where interactions between shock waves and vortical flows play a significant role \citep{Liu2022hybridDNN, kohl2023turbulent}. 
As previously mentioned, the transonic flow over an aerofoil may exhibit two distinct frequency components, depending on the free stream Mach number \citep{Bouhadji2003a}. The presence of these dual frequencies and their interaction mechanism present a significant challenge for neural network prediction models. Therefore, a specialized approach is required to capture and predict the complex interaction between these high and low-frequency phenomena.

%opened up new possibilities for performing data-driven predictive modeling of transonic flows \citep{Chen2023towards, Deng2023pof}. 

%Our main contribution is showcasing how a deep learning-based coarse model effectively captures the intricate dynamics of high-resolution unsteady transonic flows. Remarkably, the trained model can predict the evolution of the entire flowfield surrounding an aerofoil, even in scenarios with unseen free-stream Mach numbers and initial time-mean flowfields. Once trained, its adaptability extends to handling diverse tasks with multiple objectives. Furthermore, we delve into examining if the model truly captures the essence of the physical system's dynamics by scrutinizing global instability. Given the model's differentiability, we seamlessly integrate it with a linearized operator-based global instability analysis. This fusion bridges the gap between deep neural network models and traditional modal analysis, offering fresh insights into transonic flow dynamics. Our groundbreaking approach not only opens up new vistas in flowfield diagnostics but also paves the way for the advancement of interpretable neural network models.

Our main contribution is to showcase that a deep learning-based model can capture the intricate dynamics of high-resolution unsteady transonic flows. Remarkably, the trained model can predict the evolution of the entire flowfield surrounding an aerofoil, in scenarios with unseen free-stream Mach numbers and even with an initial time-mean flowfields. Once trained, its adaptability allows for handling diverse tasks with multiple objectives. 
Furthermore, we show that the resulting model really learned the essence of the underlying physical system by scrutinizing global instability. Compared to others based on a Koopman framework \cite[]{Mauroy2020} or a dynamical mode decomposition method \cite[]{herrmann_baddoo_semaan_brunton_mckeon_2021}, we show for the first time that a global instability analysis can be derived for neural networks with a linearized operator-based approach. In contrast to traditional approaches, the differentiable nature of neural network representations provides a natural avenue for evaluating fundamental physical properties of learned dynamics with 
global instability analysis.
%resolvent analysis.
%
This fusion bridges the gap between deep neural network models and traditional modal analysis, offering fresh insights into transonic flow dynamics. 
%Our groundbreaking approach not only opens up new vistas in flowfield diagnostics but also paves the way for the advancement of interpretable neural network models.
% oversell
Our innovative approach not only offers new perspectives in flowfield diagnostics but also lays the groundwork for
improving the interpretability of neural network models.
%the development of interpretable neural network models.}

%Our primary contribution is to demonstrate the deep learning-based coarse model learns the dynamics of the unsteady transonic flow at high resolution and is capable of predicting evolution of the whole flowfield around an aerofoil under unseen conditions, such as an unseen free-stream Mach numbers, 
%and
%an initial time-mean flowfields.
%and an initial Reynolds averaged Navier-Stokes result.
%Once trained, the model can be used in different tasks with multiple objectives.
%The other contribution of the present study is to examine if the model learns the dynamics of the physical system by looking into global instability.
%As the model is differentiable, we integrate it 
%with linearized operator-based global instability analysis,
%which establishes the connection between deep neural network models 
%and classical modal analysis.
%%%%% Global instability and resolvent analysis %%%
%Our proposed methodology not only offers a unique perspective on transonic flow dynamics but also presents a promising avenue for flowfield diagnostics and the development of interpretable neural network models. 

This paper is organized as follows. The physical system and data generation are briefly presented in \S \ref{sec:methodology}. The neural network architecture and training procedure will be described in \S \ref{sec:neural_network}. The detailed experiments and results are then given in \S \ref{sec:results_discussion} and concluding remarks in \S \ref{sec:concluding_remarks}.

\section{Physical system and data generation}\label{sec:methodology}
\subsection{Physical system}
We consider the transonic flow over a NACA0012 aerofoil at a low Reynolds number, i.e. $15000$. The governing equations of the physical system are compressible Navier-Stokes equations, which are non-dimensionalized with the freestream
variables (i.e., the density $\rho_{\infty}$, speed of sound $a_{\infty}$, 
%velocity magnitude $||\mathbf{v}||_{\infty}$,
and the chord length of the aerofoil $c$). The conservative form about density $\rho$, momentum $\rho u_{\mathrm{i}}$ and total energy $\rho E$ can be expressed in tensor notation as
\begin{align*}
    %\begin{split}
\frac{\partial \rho}{\partial t} + \frac{\partial \rho u_\mathrm{i}}{\partial x_\mathrm{i}}&=0 \\
\frac{\partial \rho u_\mathrm{i}}{\partial t}+ \frac{\partial \rho u_\mathrm{i} u_\mathrm{j}}{\partial x_\mathrm{j}}&=-\frac{\partial p}{\partial x_\mathrm{i}} + \frac{\partial \tau_{x_{\mathrm{i}} x_{\mathrm{j}}}}{\partial x_\mathrm{j}} \\
\frac{\partial \rho E}{\partial t}+\frac{\partial (\rho E + p)u_\mathrm{i}}{\partial x_\mathrm{i}} &= \frac{\partial (-q_\mathrm{i} +u_\mathrm{j} \tau_{x_{\mathrm{i}} x_{\mathrm{j}}} )}{\partial x_\mathrm{i}}
   % \end{split}
\end{align*}%
where the shear stress and heat flux terms are defined as
\begin{equation*}
\tau_{x_{\mathrm{i}} x_{\mathrm{j}}}=\mu\frac{M_{\infty}}{Re_{\infty}}[(\frac{\partial u_\mathrm{i}}{\partial x_\mathrm{j}}+\frac{\partial u_\mathrm{j}}{\partial x_\mathrm{i}})-\frac{2}{3}\frac{\partial u_{\mathrm{k}}}{\partial x_{\mathrm{k}}}\delta_{\mathrm{i}\mathrm{j}}]
\end{equation*} 
and 
\begin{equation*}
q_{x_{\mathrm{i}}}=-\frac{\mu}{Pr}\frac{M_{\infty}}{Re_{\infty}(\gamma-1)}\frac{\partial \Theta}{\partial x_{\mathrm{i}}}.
\end{equation*}
Here, Reynolds number is defined as $Re_{\infty}=\rho_{\infty}\sqrt{u_{\infty}^2+v_{\infty}^2}c/\mu_{\infty}$; $\gamma$ is the ratio of specific heats, 1.4 for air; the laminar viscosity $\mu$ is obtained by Sutherland's law (the function of temperature); laminar Prandtl number is constant, i.e.
$Pr=0.72$.
The relation between pressure $p$ and total energy $E$ is given by
\begin{equation*}
p=(\gamma-1)\bigl[\rho E-\frac{1}{2}\rho u_{\mathrm{i}}u_{\mathrm{i}}\bigr].
\end{equation*}
Note also that from the equation of state for a perfect gas, we have $p=\rho a^2/\gamma$ and temperature $\Theta=a^2$.
No turbulence model is employed as we perform 2D high-resolution quasi-direct numerical simulations. 

\begin{figure}
\centering
\begin{subfigure}{.5\textwidth}
\centering
\hspace{-4em} % Adjust the value as needed to move the figure left
\includegraphics[width=\linewidth]{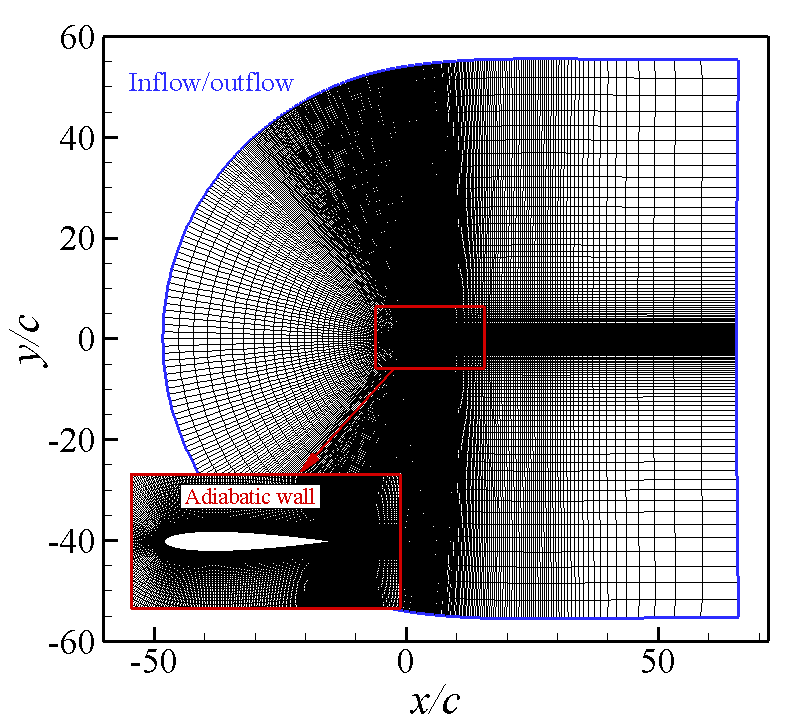} 
\caption{}
\end{subfigure}
\begin{subfigure}{.6\textwidth}
\centering
\includegraphics[width=\linewidth]{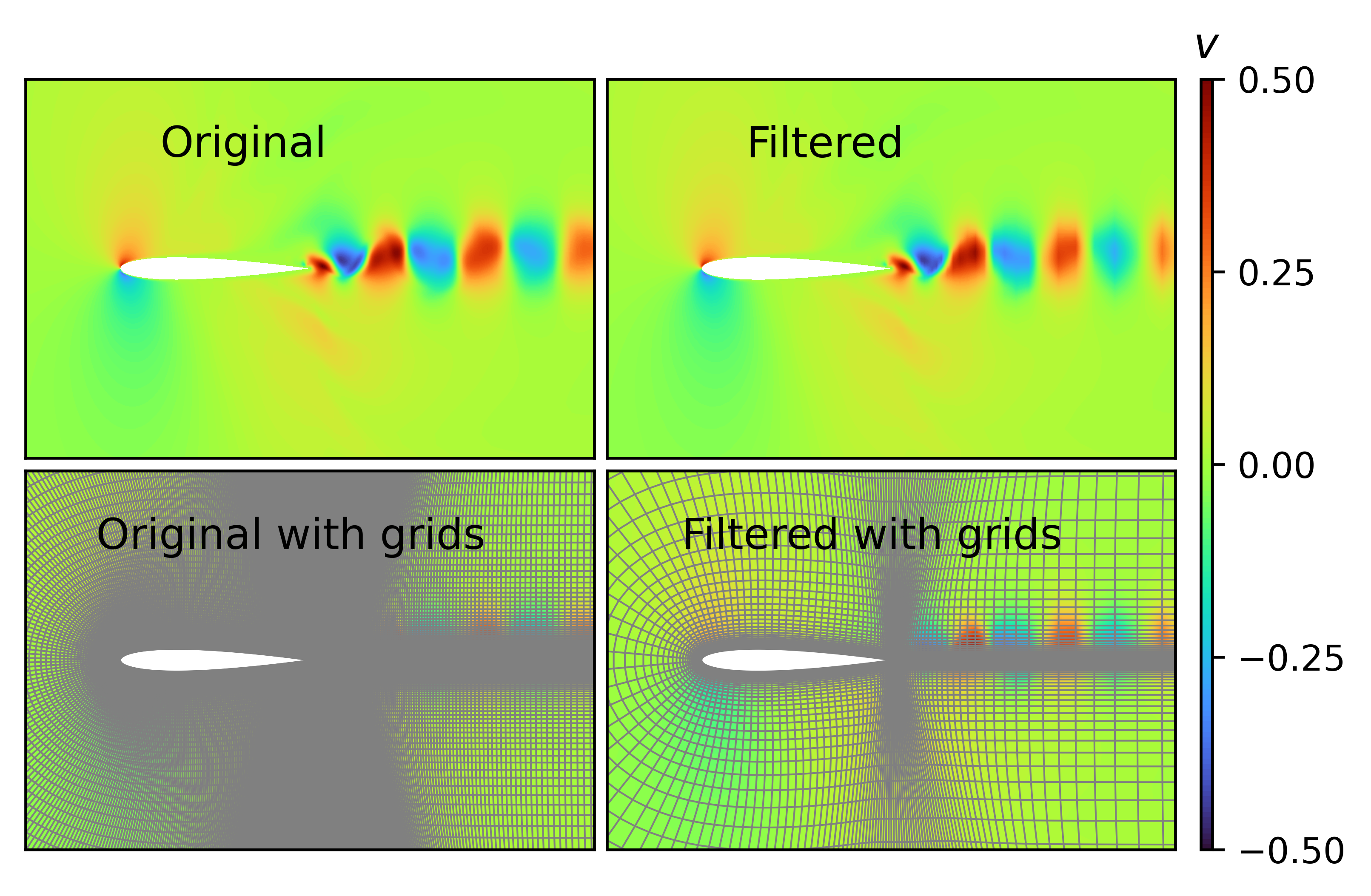} 
\caption{}
\end{subfigure}
    \caption{(a) Computational mesh with $1024\times128$ grid cells; (b) The contour of the y-component velocity at $M_{\infty}=0.825$ on the fine grids on the left column v.s. filtered one on coarse grids ($256\times64$) on the right column. The bottom shows the fields with the corresponding meshes.}
    \label{fig:computational_mesh}
\end{figure}

The finite-volume method numerically solves the equations using the open-source code CFL3D \citep{Rumsey1997, Rumsey2010}. As depicted in figure \ref{fig:computational_mesh}(\textit{a}), the mesh resolution $1024\times128$ is kept the same for cases, i.e. 128 grid cells in the wall-normal direction, 320 grid cells in the wake, and 384 grid cells around the aerofoil surface. 
The convective terms are discretized with the third-order upwind scheme and viscous terms with a second-order central difference.  
An inflow/outflow boundary based on one-dimensional Riemann invariant is imposed at about 50\emph{c} away from the aerofoil in the (\emph{x}, \emph{y}) plane. The grid stretching is employed to provide higher resolution near the surface and in the wake region, and the minimal wall-normal grid spacing is $6\times10^{-4}$ to ensure $y^+_n<1.0$. A no-slip adiabatic wall boundary condition is applied on the aerofoil surface. 
The non-dimensional time step is $0.008c/{U_{\infty}}$. 
% dt_a = dt/(c/a)=dt*a/c; dt_u = dt/(c/U)=dt*U/c
% dt_U/dt_a = M
% dt_U = 0.008 --> dt= 0.008*c/U

\begin{figure}
    \centering
    \includegraphics[width=9cm]{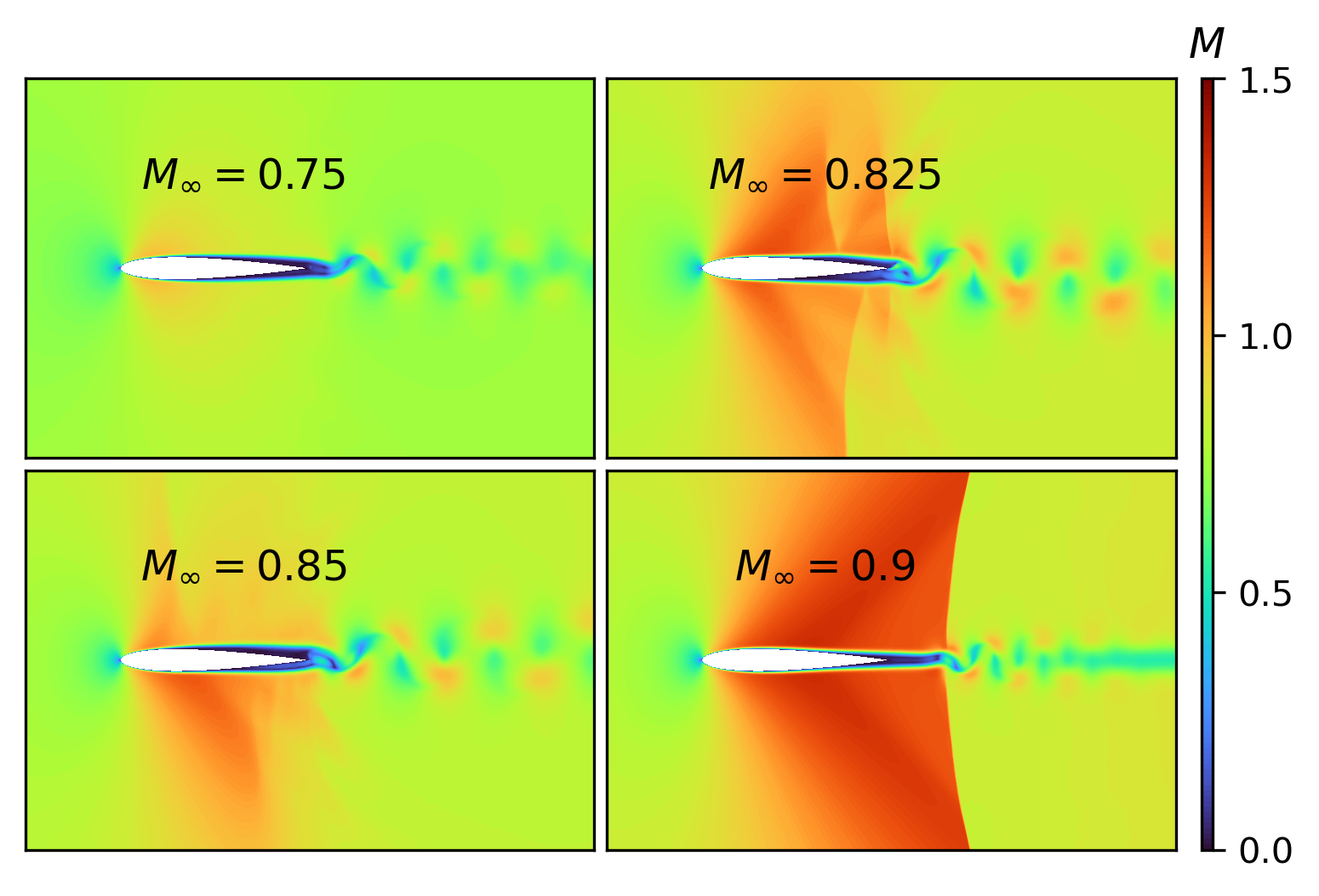}
    \caption{Typical flowfields in the training set.}
    \label{fig:typical_flowfields}
\end{figure}

\begin{figure}
\centering
\begin{subfigure}{.45\textwidth}
\centering
\includegraphics[width=\linewidth]{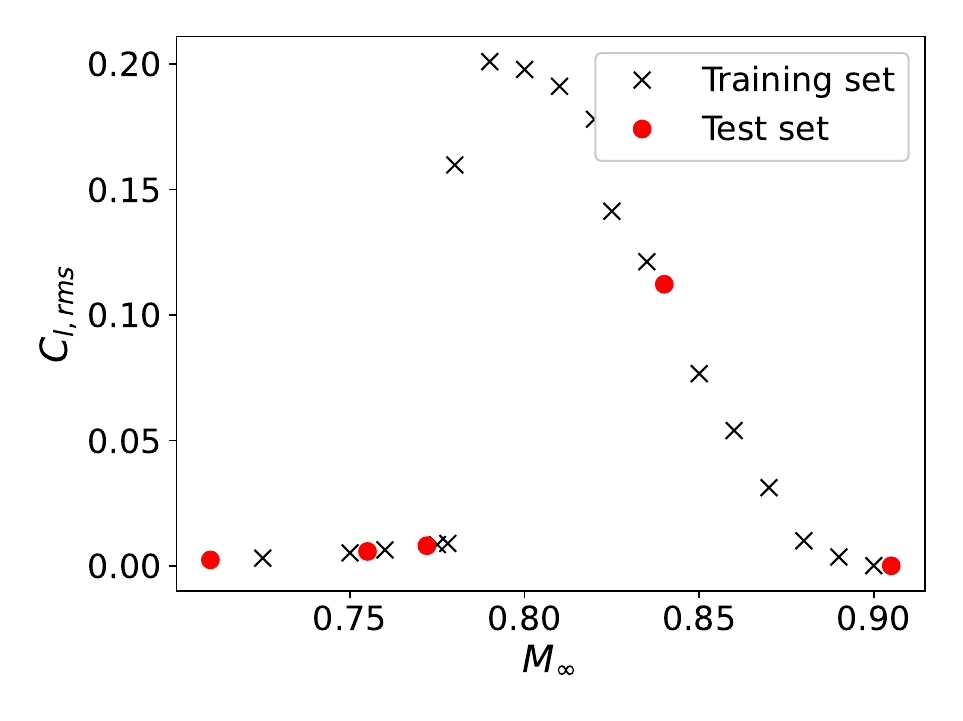}
\caption{}
\end{subfigure}
\begin{subfigure}{.45\textwidth}
\centering
\includegraphics[width=\linewidth]{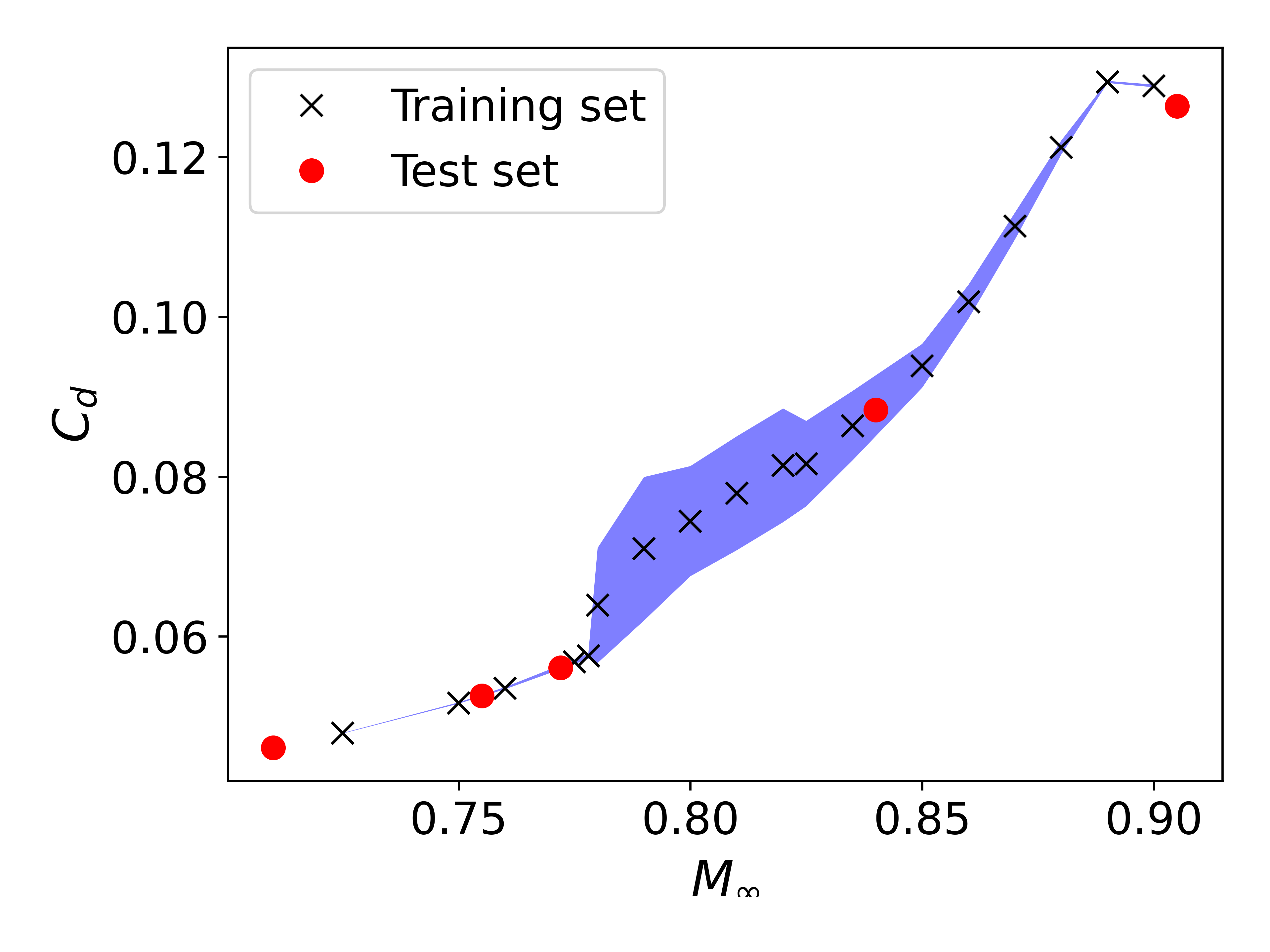}
\caption{}
\end{subfigure}
\caption{(a) Root mean square values of lift coefficients and (b) Statistically mean drag coefficients; shaded regions represent $\pm3$ standard deviations.}
    \label{fig:lift_and_drag}
\end{figure}

\subsection{Data generation}
In the transonic regime, airflow behavior becomes more complex due to the formation of shock waves and supersonic and subsonic flow areas on the aerofoil surfaces. To cover all possible flow regimes of interest, the samples in the training dataset are generated at 18 different Mach numbers (i.e., $M_{\infty}$=0.725, 0.75, 0.76, 0.775, 0.778, 0.78, 0.79, 0.8, 0.81, 0.82, 0.825, 0.835, 0.85, 0.86, 0.87, 0.88, 0.89, and 0.9).

As shown in figure \ref{fig:typical_flowfields}, at $M_{\infty}=0.75$, the primary source of unsteadiness in the flowfield stems from high-frequency vortex shedding. When the free-stream Mach number approaches 0.8, a series of compression waves coalesce to form a strong shock wave, and the flow structures are dominated by alternately moving shock waves along the upper and lower sides of the aerofoil (e.g. $M=0.825$ and $M=0.85$ in figure \ref{fig:typical_flowfields}). In this regime, the flowfield exhibits a combination of high-frequency vortex shedding and low-frequency unsteadiness attributed to shock-wave dynamics, leading to significant fluctuations in lift and drag coefficients, as shown in the figure \ref{fig:lift_and_drag}. When the Mach number exceeds $0.89$, the intense shock waves stabilize on both surfaces, and unsteadiness is predominantly driven by the shedding of vortices in the wake.

To evaluate the learned model, we choose five representative conditions for the test set: $M_{\infty}=0.71$, $M_{\infty}=0.755$, and $M_{\infty}=0.772$ are in the vortex shedding regime; %$M_{\infty}=0.783$ 
$M_{\infty}=0.84$ case is in the shock oscillation regime; $M_{\infty}=0.905$ is the case when the shockwave is stable. The lift and drag coefficients of the test cases are marked as red symbols in figure \ref{fig:lift_and_drag}.

Following the completion of the transient phase, snapshots are stored at intervals of every four simulation steps (i.e., $dt_{sampling}=0.032c/U_{\infty}$ and  $f_{Nyquist}=15.625U_{\infty}/c$). As the flowfield variables are defined on grid cells, we downsample them by 4$\times$ and 2$\times$ in the circumferential and wall-normal directions with spatial filtering. Specifically, the density and pressure, as well as the grid cell center coordinates, are subject to cell-area-weighted averaging, while the velocities undergo mass-weighted filtering, known as Favre filtering, preserving the grid's topological structure and maintaining a one-to-one block interface in the wake region.
The typical filtered field variables are shown in figure\ref{fig:computational_mesh}(\textit{b}). In each case (or sequence), there are 400 to 1000 snapshots, which capture at least two periods of low-frequency motion in the flowfield.

Through the coordinate transformation, variables in the physical space can be uniquely mapped into the computational space and back \cite[]{Gordon1973constructionCurvilinear}.
Hence, we define a unit cube of curvilinear coordinates as the canonical space in which neural networks operate \cite[]{Chen2023towards}. 
As our training focuses on a single aerofoil and computational mesh, the geometric information, in the form of a transformation matrix and airfoil profile, is not included as part of the neural network's input. 
%However, it's important to note that since we use only one set of mesh and one aerofoil profile, the geometric information, such as the transformation matrix and airfoil profiles, is not included as part of the neural network's input. 

\section{Predictive modelling with deep learning}\label{sec:neural_network}
\subsection{Learning task}
We frame the learning task as the discovery of a function $f_{\theta}$ (with weights $\theta$) that can map the input state variables $\mathbf{q}_{t}$, given the condition $M_{\infty}$, to an approximation of the subsequent state $\mathbf{q}_{t+\Delta t}$, 
\begin{equation}\label{eqn:nn}
    \mathbf{q}_{t+\Delta t}=f_{\theta}(\mathbf{q}_{t}, M_{\infty}).
\end{equation}
Here, $\mathbf{q}$ represents the four flowfield variables, including density, velocities in the $x$ and $y$ directions, and pressure. The learned model is applied for a sequence of $m$ steps, generating a trajectory of states $(\mathbf{q}_{t_0}, \dots, \mathbf{q}_{t_m})$, and training is aimed at effectively capturing the dynamic evolution of a dataset of ground truth trajectories.$(\mathbf{\tilde{q}}_{t_0}\dots\mathbf{\tilde{q}}_{t_m})$.

\subsection{Neural Network Architectures}
%We implemented Attention 
As architecture for the neural networks we employ an attention-based U-Net  \citep{oktay2018attention}. It consists of three encoder blocks, each progressively capturing features from the input image through convolution with Group Normalization and ReLU activation, and downsampling. Following the encoder blocks, there's a bottleneck block for information compression with a higher number of channels. Subsequently, the architecture includes three decoder blocks, which use skip connections to integrate features from both the bottleneck and corresponding encoder blocks during the upscaling process. These decoder blocks gradually reduce the number of channels, culminating in a $1\times1$ convolutional output layer that generates pixel-wise predictions. 
In contrast to the original implementation, our current study places a specific emphasis on the c-type mesh topology. As a result, we've incorporated custom paddings for all convolution layers and bilinear upsampling operations, achieved by exchanging the 1-1 block interface values within the mesh interface. This approach allows us to preserve the c-type mesh topological structure throughout our feature maps.

We train networks of varying sizes by adjusting the number of features, as indicated in table \ref{tab:aero-arch}. The details are provided in the appendix figure \ref{fig-App:details_about_attentionUnet}.

\begin{table}
    \centering
    \begin{tabular}{l| r c l }
        \toprule
        \textit{Model name} &   \textit{\# of Parameters}  & \textit{\# of Features in encoder  \& bottleneck} \\
        \midrule
%        UNet, 0.13m &  128756  & [8, 16, 32] [64] \\ 
        0.51M &  511332  & [16, 32, 64] [128] \\ 
        2.04M & 2037956  & [32, 64, 128] [256] \\
        8.14M & 8137092  & [64, 128, 256] [512] \\
        \bottomrule
    \end{tabular}
    \caption{Network architectures and parameter counts}
    \label{tab:aero-arch}
\end{table}

\subsection{Rollout training}
Prediction operates on the same network parameter (i.e. $\theta$) from equation \ref{eqn:nn}. The total loss over the unrolled trajectory consisting of $s$ steps is 
\begin{equation}
    \mathcal{L} = \sum_{i=1}^s  \mathcal{L}(\tilde{\mathbf{q}}_{i+s\Delta t}, f_\theta^s(\mathbf{q}_i)) = \sum_{i=1}^s  \mathcal{L}^s.
\end{equation}
Here, $\mathcal{L}^s$ represents the loss evaluated at a specific step $s$. 
As the model is differentiable, we can calculate the full optimization gradients by propagating gradients through predictions. The gradients are thus evaluated as 
\begin{equation}
    \frac{\partial \mathcal{L}^s}{\partial \theta} = \sum_{B=1}^s \bigg[
    \frac{\partial \mathcal{L}^s}{\partial f_\theta^s} 
    \bigg ( \prod_{b=1}^{s-B}
    \frac{\partial f_\theta^{s-b+1}}{\partial f_\theta^{s-b}} 
    \bigg)
    \frac{\partial f_\theta^B}{\partial \theta}\bigg].
\end{equation}

The model is trained with Adam and a mini-batch size of 4. The group normalization technique employs four groups within the convolutional layers. To improve generalization on unseen data and enhance prediction robustness, we introduce 0.5\% Gaussian noise into the input data \citep{bishop:2006:PRML, kohl2023turbulent}.
Following \cite{um2020sol}, the training curriculum gradually increases the number of unrolled steps, commencing with $s=1$ and subsequently progressing to $s=4$, and then $s=9$, followed by $s=14$. In each, a learning rate of $6\times10^{-4}$ is applied for the initial 250k iterations, after which it exponentially decays to $6\times10^{-5}$ for the rest 250k. 
Thus, a total of 2 million iterations are used in the whole curriculum. 
Our observations indicate that the marginal benefits diminish significantly beyond 14 roll-out steps in the current context. Hence, we choose the model trained with $s=14$ for our analyses.

\section{Results and discussion}\label{sec:results_discussion}
\subsection{Inference from a previously unseen condition}
In the following, we discuss the overall performance of the learned models and the predicted aerodynamic characteristics. We investigate test cases with previously unseen Mach numbers: $M_{\infty}=0.71$, $M_{\infty}=0.755$, $M_{\infty}=0.772$, %$M_{\infty}=0.783$, 
$M_{\infty}=0.84$ and $M_{\infty}=0.905$, corresponding to three shock-free cases,
one shock-oscillation case
and one stabilized shock-wave case, respectively. Given the Mach number range of the training data, cases of $M_{\infty}=0.71$ and $M_{\infty}=0.905$ will be considered as ``extrapolation tests'', the other three will be treated as ``interpolation tests''.

\subsubsection{Overall evaluation}

\begin{figure}
    \centering
\begin{subfigure}{.45\textwidth}
\centering
\includegraphics[width=\linewidth]{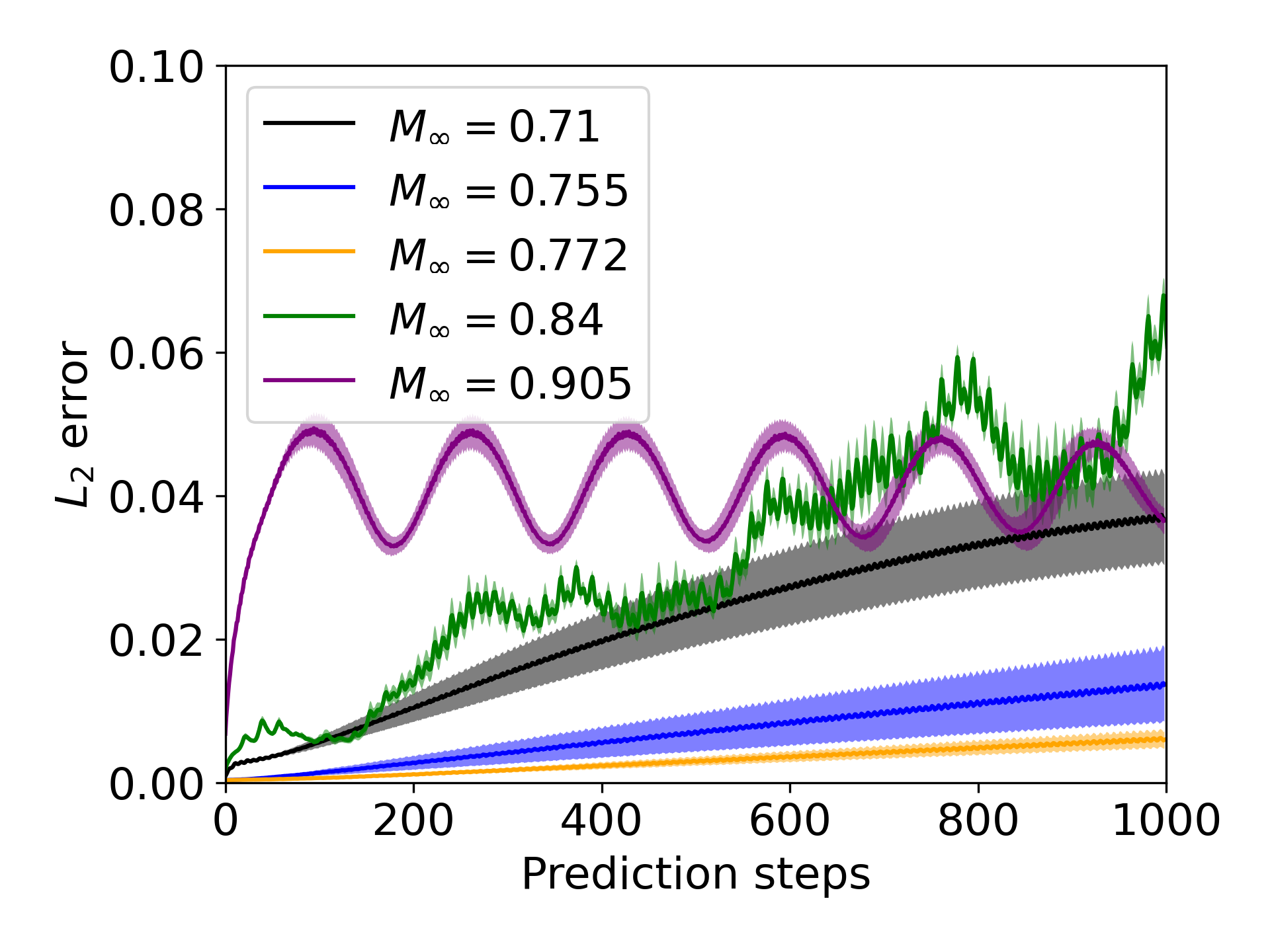}
\caption{Loss history over 999 prediction steps.}
\end{subfigure}
\begin{subfigure}{.45\textwidth}
\centering
\includegraphics[width=\linewidth]{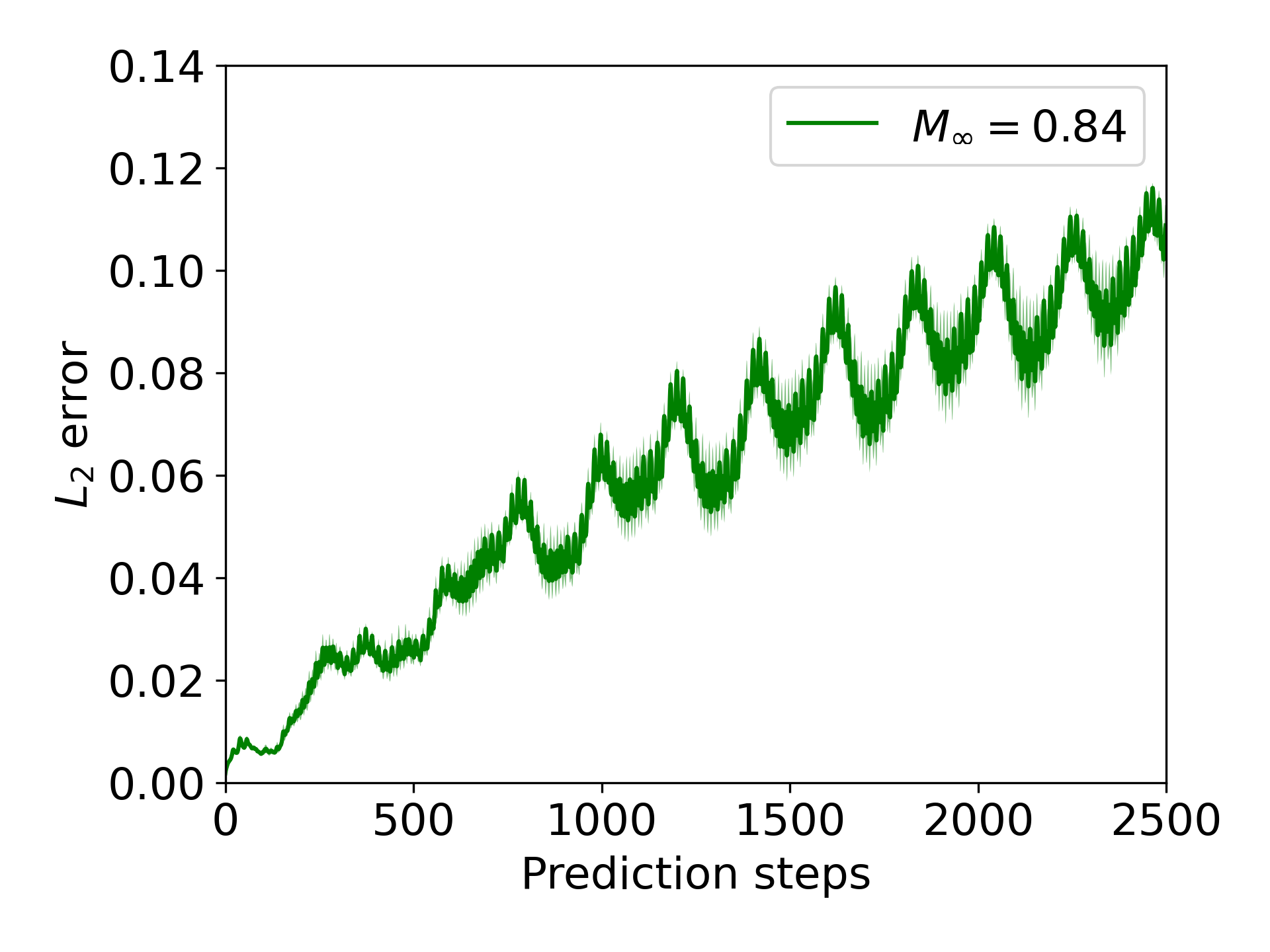}
\caption{Loss history over 2499 prediction steps.}
\end{subfigure}

    \caption{Long-term $L_2$ errors measured using five large-network-size models (8.14M), each initialized with distinct random seeds.}
    \label{fig:loss_prediction_steps}
\end{figure}

\begin{table}
    \centering
    \begin{tabular}{l| c | c c c }
        \toprule
        Case & Model name &  $N=99$  & $N=999$  & $N=2499$ \\
        \midrule
             &0.51M     & 0.00146$\pm$0.000497 &  0.0138$\pm$0.0102  & / \\
        $M=0.755$ &2.04M &0.000956$\pm$0.000272    & 0.00937$\pm$0.00390 & /  \\ 
         & 8.14M & \textbf{0.000813}$\pm$0.000304    & \textbf{0.00695}$\pm$0.00517   & / \\
          \hline
             &  0.51M       &  0.00144$\pm$0.000462   & 0.0117$\pm$0.00794   & / \\
        $M=0.772$ & 2.04M & 0.000743$\pm$ 0.000168   & 0.00451$\pm$0.00394 & / \\
         & 8.14M & \textbf{0.000511}$ \pm 6.91\times10^{-5}$    &\textbf{0.00305}$\pm$0.00117    & / \\
                  \hline
          &  0.51M    &  \textbf{0.00545}$\pm$0.000536 & \textbf{0.0167}$\pm$0.00465  & \textbf{0.0325}$\pm$0.00978 \\
          % 0.016705967 +/- 0.0046464703
        $M=0.84$ & 2.04M & 0.00578$\pm$0.000478   & 0.0229$\pm$0.00663  &  0.0470$\pm$0.0134 \\
       & 8.14M  & 0.00629 $\pm$0.000668 & 0.0304$\pm$0.00386  & 0.0609$\pm$0.00619 \\
        \hline
        \hline
        & 0.51M & 0.00369$\pm$0.000774 & \textbf{0.0186}$\pm$0.00686& / \\
        $M=0.71$ & 2.04M & \textbf{0.00355}$\pm$ 0.000435   &0.0191$\pm$0.00586 & / \\ 
        & 8.14M &  0.00371$\pm$0.000686   & 0.0221$\pm$0.00818  & /\\
        \hline
        & 0.51M & 0.0412$\pm$0.00564    & 0.0575$\pm$0.0202  & /\\
        $M=0.905$& 2.04M & 0.0389$\pm$0.00239   & 0.0432$\pm$0.00568 & / \\
        & 8.14M   & \textbf{0.0374}$\pm$0.00255    &  \textbf{0.0411}$\pm$0.00445  & /\\
%        $M=0.71$ &    &0 0.019064656$\pm$0.0058625867\\ 
%        $M=0.755$ &    & 0.009373766$\pm$0.0038950795  \\ 
%        $M=0.772$  &    & 0.0045081656$\pm$0.0039439723 \\
%        $M=0.84$ &    & 0.0228624$\pm$0.006632131  &  0.047046416$\pm$0.013362877 \\ 
%        $M=0.905$ &    & 0.04322267$\pm$0.005685139 \\

% large size:
%3 : Mach=0.84  <-------- Only this makes sense 2500

%  ----------------------999 ---------------
%1 : Mach=0.755 
%2 : Mach=0.772 
%3 : Mach=0.84 

%0 : Mach=0.71 
%4 : Mach=0.905

% -----------------------99 -----------------
%2 : Mach=0.772 
%3 : Mach=0.84 

%0 : Mach=0.71 
%4 : Mach=0.905 

        \bottomrule
    \end{tabular}
    \caption{Accumulated $L_2$ errors for $N$ prediction steps.}
    \label{tab:loss-accumulated}
\end{table}

Due to the different random seeds used for initializing the training of neural networks, the resulting models exhibit certain variations in predictions. To address this, we trained five models and calculated the average prediction error at a given prediction step, as shown in figure \ref{fig:loss_prediction_steps}. The shade areas represent the $\pm0.5$ standard deviation. Figure \ref{fig:loss_prediction_steps}(a) gives the loss history over 999 prediction steps. At free-stream Mach numbers of 0.71, 0.755, and 0.772, since there are no complex shock or compression wave motions, the flow structure is primarily characterized by periodic vortex shedding. Consequently, although the error accumulates with the increase in prediction steps, the overall errors remain small. Notably, the case at Mach 0.71 is an extrapolation test, which explains its larger error compared to the other two.
We observe a notable difference in error behavior at Mach number 0.905 compared to the other three. Here, the error becomes significant early on but does not display a clear growth trend. 
This phenomenon can be attributed to the considerable deviation of the flow regime from the majority of training samples: a stable shock wave system with moderate interaction between shock waves and vortices. While phase deviations are present, their cumulative effect over time remains relatively modest.
%Looking at the other extrapolation test at a free-stream Mach number of 0.905, we observe that as the shock wave stabilizes and the vortex shedding is pulled downstream, interacting with the shockwaves, the flow regime differs significantly from the majority of samples in the training set. Consequently, the accumulated losses exhibit larger values.

At Mach number 0.84, the moving shock wave over the aerofoil surface exhibits significant unsteadiness and distortions as it interacts with the boundary layers. 
As a result, the errors observed are substantially higher than those in other scenarios shown in figure \ref{fig:loss_prediction_steps}(a). 
We are particularly intrigued by this case, and thus we have conducted extended prediction steps (2499 steps, equivalent to 9996 simulation steps). As shown in figure \ref{fig:loss_prediction_steps}(b), despite displaying an upward trend, the prediction remains stable.

As shown in table \ref{tab:loss-accumulated}, we calculate the mean $L_2$ errors and standard deviations for the first $N$ prediction steps, where $N=99$, $N=999$ and $N=2499$ (only for $M_{\infty}=0.84$), respectively. The bold texts show the highest error or standard deviations. The error values and standard deviations further illustrate that in complex physical problems such as these, it is insufficient to rely solely on the interpolation and extrapolation expectations tested by the simple categorization of data labels (i.e. Mach numbers). One must also consider whether the statistical distribution of the test data is similar to that of the training dataset. For instance, at $M_{\infty}=0.84$, although categorized as an interpolation case, the structure and evolution of the flow field are sufficiently intricate that they have deviated from the scope of the samples within the training dataset, resulting in a pronounced error. Upon comparing the performance of three different sizes of neural networks, it is evident that the model ``8.14M '' exhibits the most optimal performance overall. While the model ``8.14M '' may not consistently outperform others in individual tests, this observation also implies that the marginal benefit of further increasing the neural network size is limited for the current task. Subsequent analyses will delve deeper into the results obtained with the model ``8.14M''.

\begin{figure}
    \centering
    \includegraphics[width=.75\textwidth]{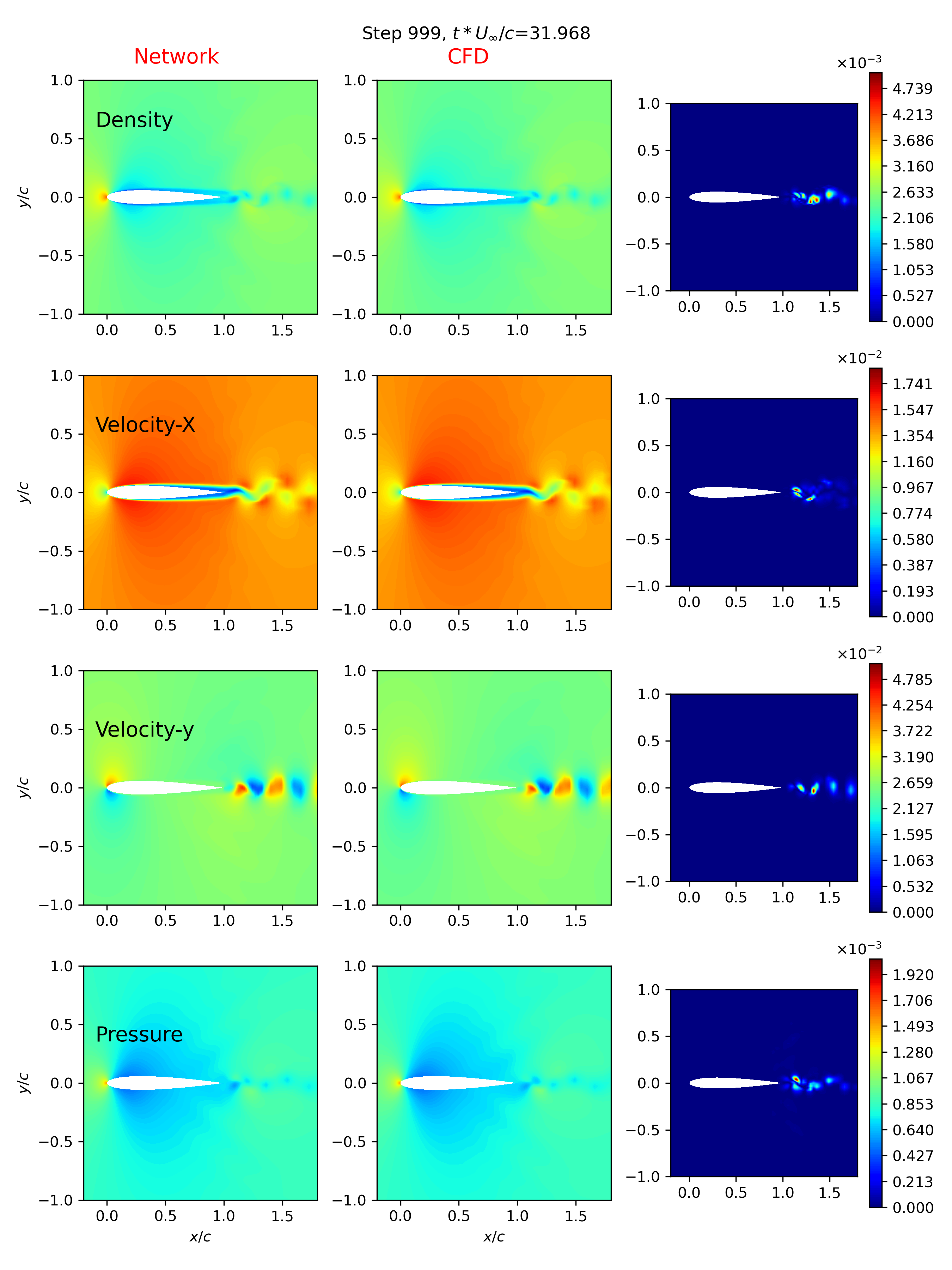}
    \caption{The flowfield from the inference at $M_{\infty}=0.755$ after 999 prediction steps.}    \label{fig:visualization_DNN_CFD_999}
\end{figure}

Now, let’s evaluate the accuracy and errors in 
%the spatial distribution of physical quantities
flowfields
predicted by the neural network model. We will mainly focus on two representative cases: the shock-free case at $M_{\infty}=0.755$ and the shock wave oscillation case at $M_{\infty}=0.84$.
Figure \ref{fig:visualization_DNN_CFD_999} compares the flow field at $M_{\infty}=0.755$ after 999 prediction steps with CFD data, showing density, two velocity components, and pressure from top to bottom, respectively. Although there are visual differences, one can still observe $O(0.01)$ deviations in the wake region. 
Figure \ref{fig:visualization_DNN_CFD_999_m0p84} illustrates the flow field at $M_{\infty}=0.84$ after 999 prediction steps, compared with CFD results. The velocity and pressure contour plots demonstrate a noticeable phase shift in the motion of shockwaves along the lower side of the aerofoil. Despite minor errors, particularly near shock waves and areas of strong compression, as well as in the wake, the overall prediction of the flow field structure is remarkably accurate. For an extended inference test, we also demonstrate that the flow field remains stable even after 2499 steps (see appendix figure \ref{fig-App:visualization_DNN_CFD_2499_m0p84}).

\subsubsection{Aerodynamic characteristics}
We discover that although the transient flow field experiences deviations due to the accumulation of errors and phase shifts, the learned model still effectively captures the statistical quantities and frequency components of the flowfield. To assess the aerodynamic characteristics, we compare the predicted results with CFD data in terms of the time-averaged pressure coefficient and its fluctuations. Additionally, we analyze the FFT spectrum of the pressure signal measured at a point close to the aerofoil trailing edge, specifically at $x/c=0.972$, $y/c=-0.00452$.

\begin{figure}
    \centering
    \includegraphics[width=.75\textwidth]{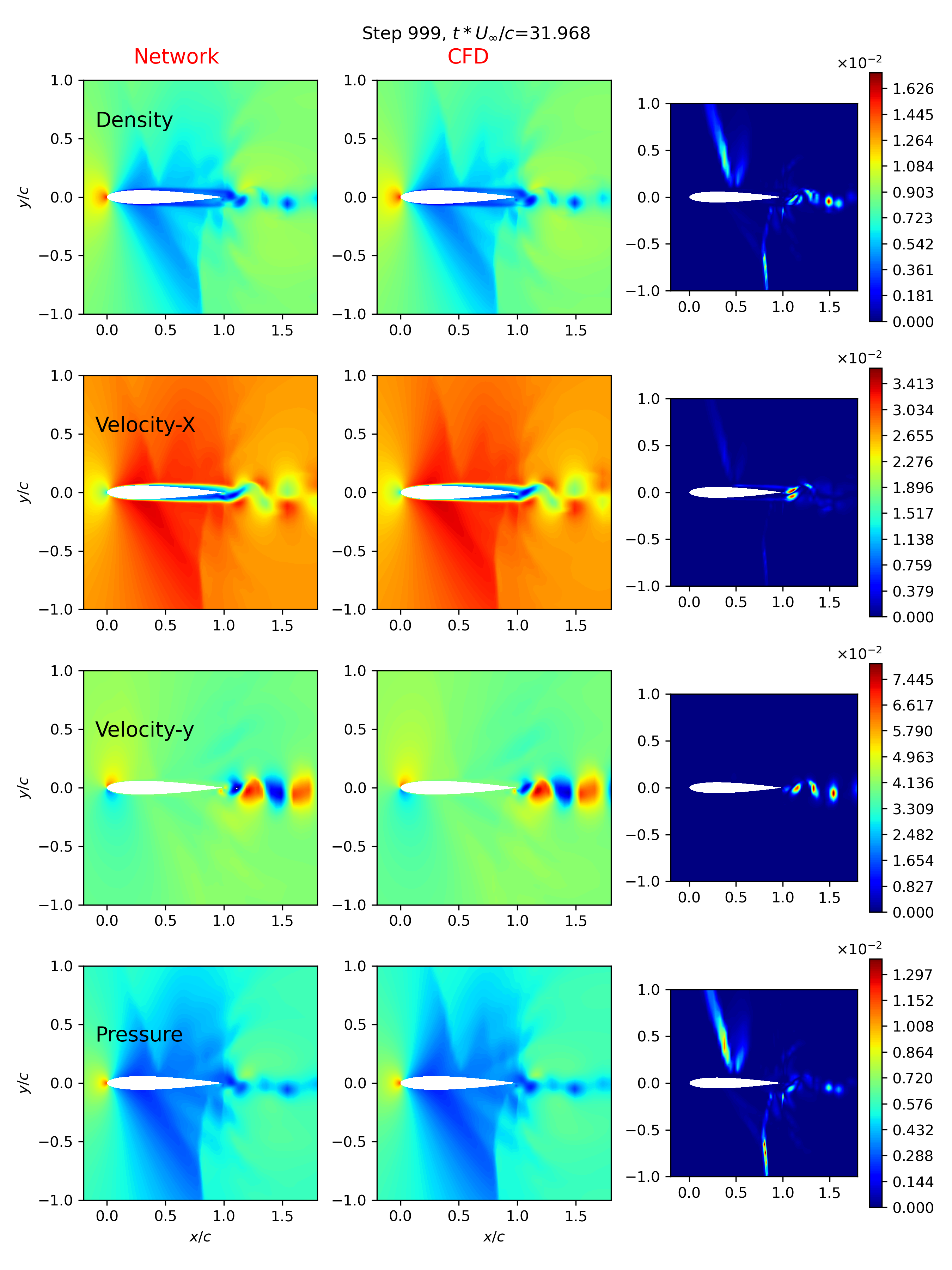}
    \caption{The flowfield from the inference at $M_{\infty}=0.84$ after 999 prediction steps.}
    \label{fig:visualization_DNN_CFD_999_m0p84}
\end{figure}

\begin{figure}
    \centering
    
\begin{subfigure}{.45\textwidth}
\centering
\includegraphics[width=\linewidth]{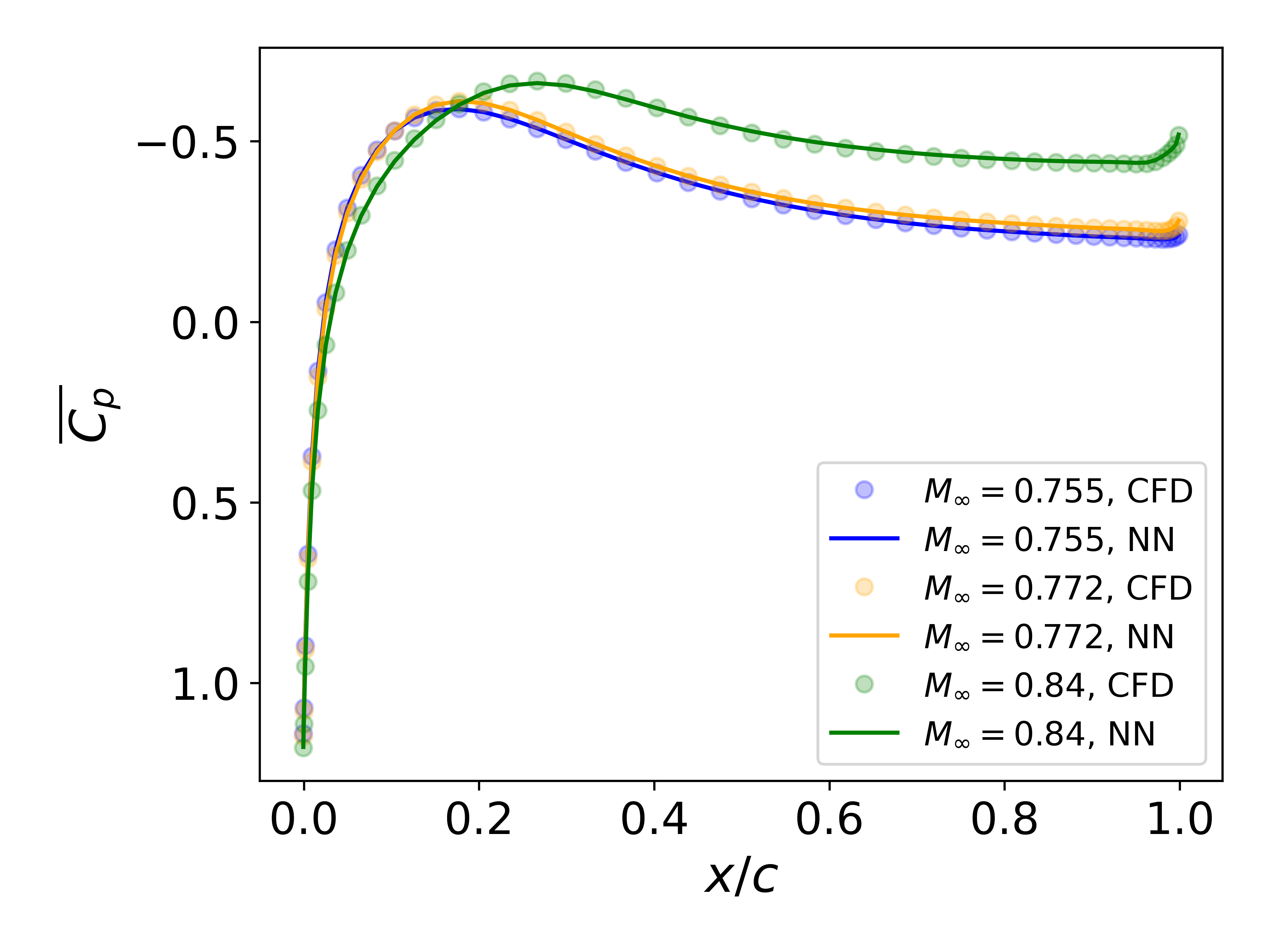}
\caption{}
\end{subfigure}
\begin{subfigure}{.45\textwidth}
\centering
\includegraphics[width=\linewidth]{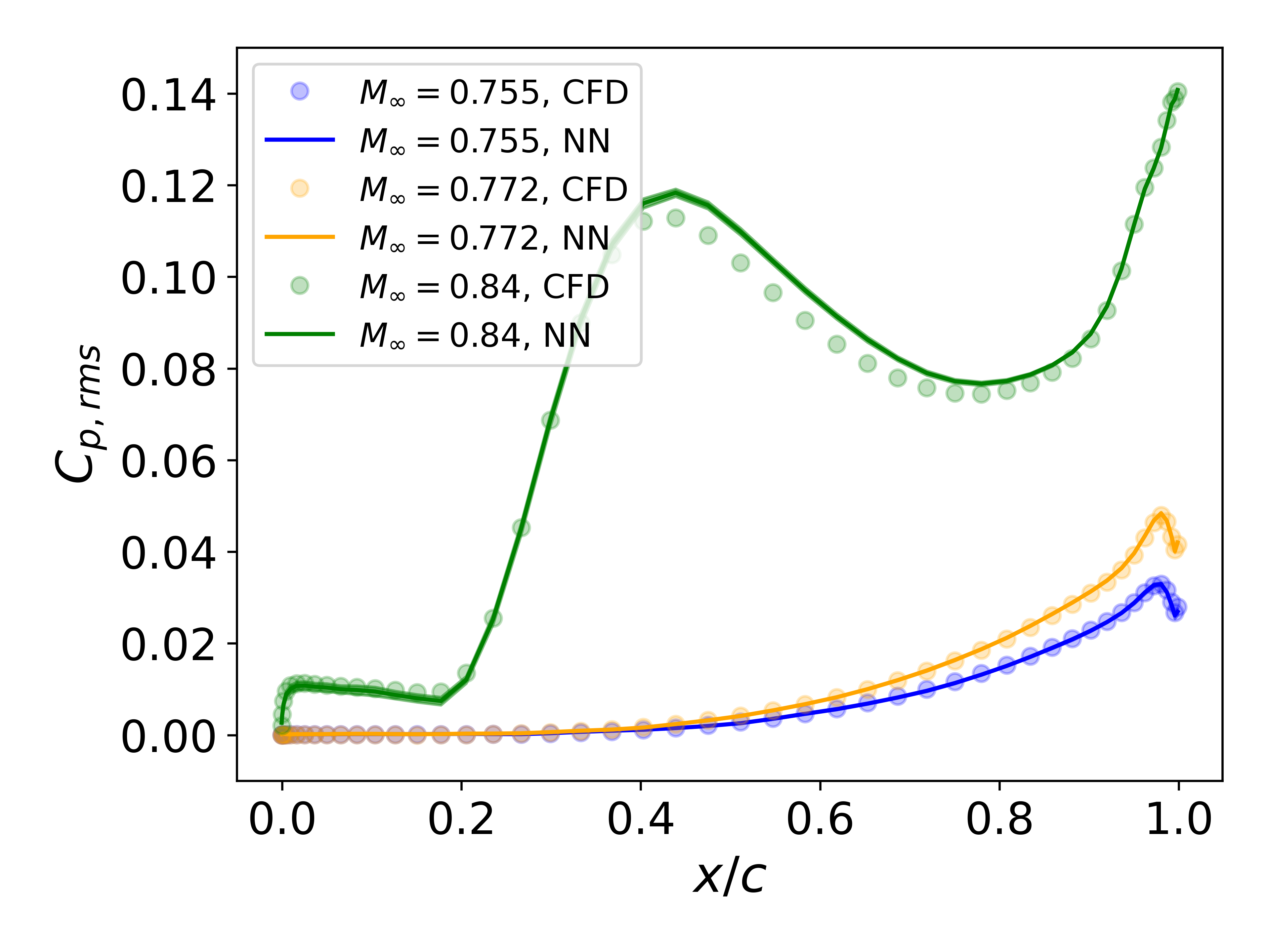}
\caption{}
\end{subfigure}

\begin{subfigure}{.45\textwidth}
\centering
\includegraphics[width=\linewidth]{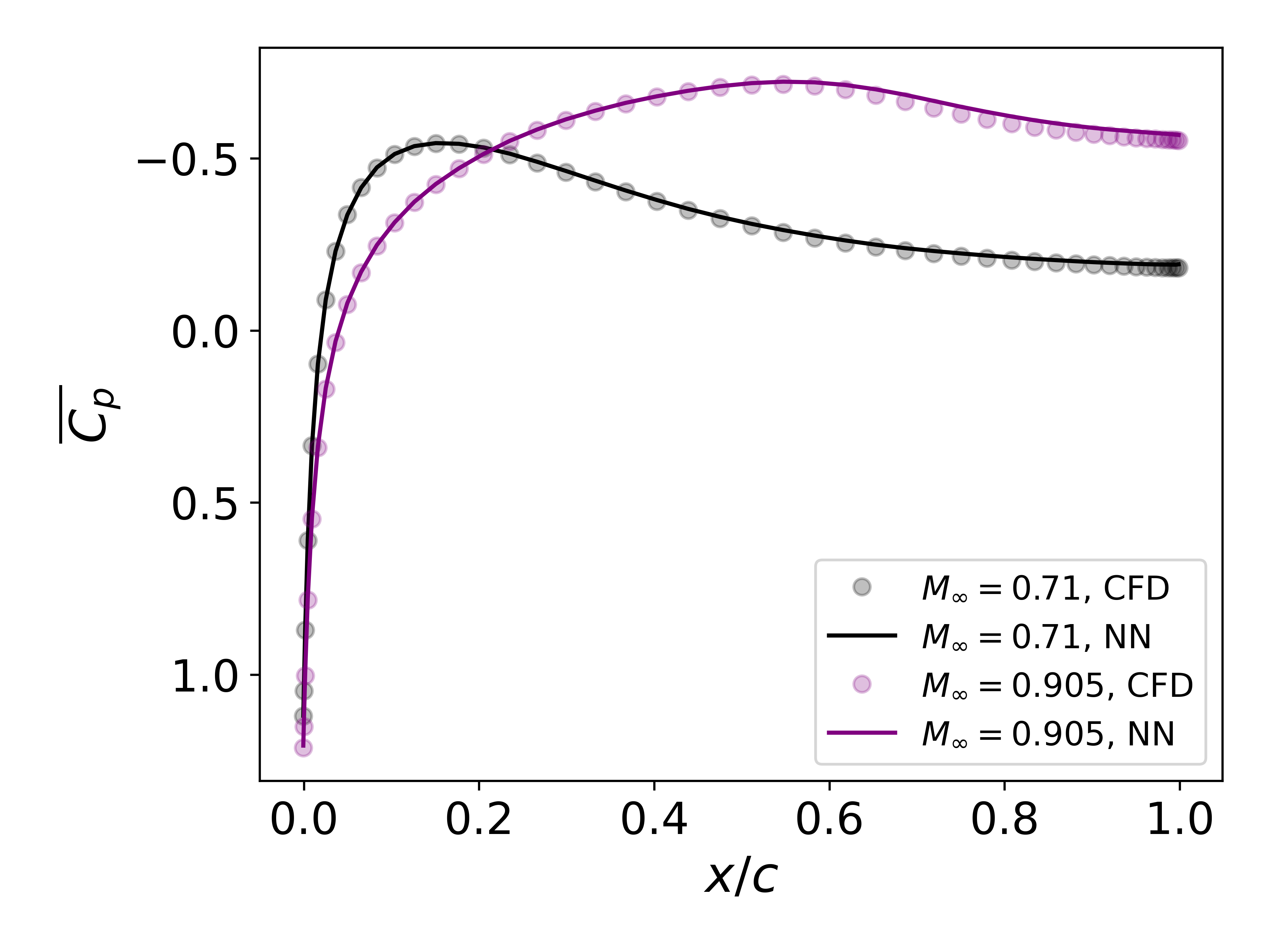}
\caption{}
\end{subfigure}
\begin{subfigure}{.45\textwidth}
\centering
\includegraphics[width=\linewidth]{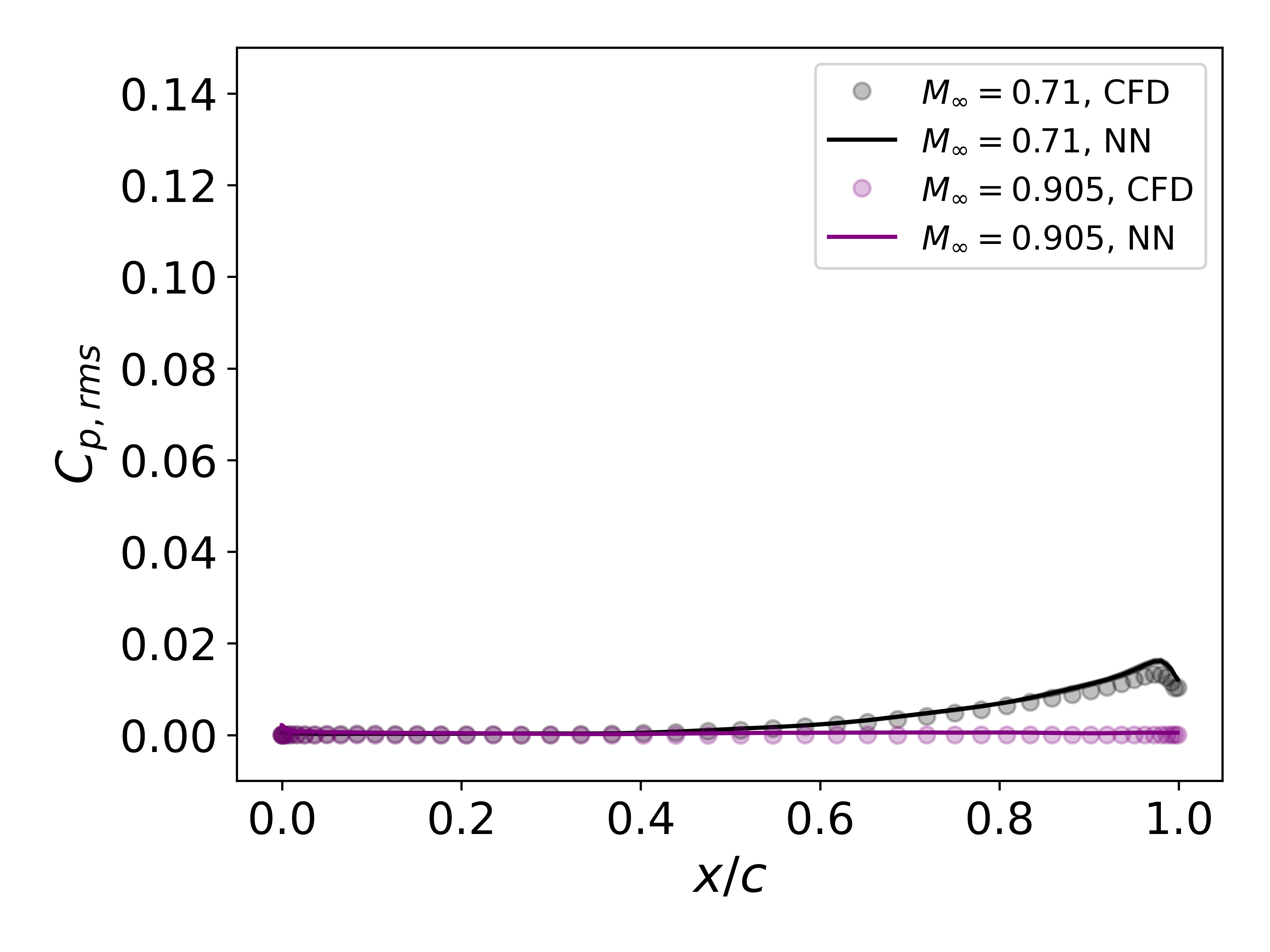}
\caption{}
\end{subfigure}

\caption{Pressure coefficients and fluctuations: (a-b) represent results from interpolation tests, while (c-d) depict outcomes from extrapolation tests. The inference is initialized with an unseen instantaneous flowfield.}
\label{fig:cp_mean_and_rms}
\end{figure}

Figure \ref{fig:cp_mean_and_rms} (a) shows the distribution of the pressure coefficient averaged over time on the surface of the airfoil, where the shading indicates the range of $0.5\sigma$ ($\sigma$, standard deviation) above and below the average value, showing the uncertainty in the prediction of learned models using different initialization weights.
First, let us consider the interpolation tests at Mach numbers 0.755, 0.772, and 0.84. Figures \ref{fig:cp_mean_and_rms} (a) and \ref{fig:cp_mean_and_rms} (b) show that the predictions are in excellent agreement with the CFD data. The shade areas indicating the uncertainty caused by random seeds are not pronounced. 
In figure \ref{fig:cp_mean_and_rms}(b), the higher values close to $x/c=1.0$ are caused by a series of compression waves that develop in the region near the trailing edge.
Note that at Mach number 0.84, the pressure fluctuation shows a peak at around $x/c=0.45$, corresponding to the mean location of the strong shock. Furthermore, the observation that the pressure fluctuation is higher than in the other two scenarios corresponds well with the complex wave systems and strong unsteady processes inherent in the flow field, as depicted in Figure \ref{fig:visualization_DNN_CFD_999_m0p84}.
The peak value around $x/c=0.4$ is attributed to the formation of a strong shock wave.

Let's further look at the extrapolation tests. 
As illustrated in Figure \ref{fig:cp_mean_and_rms}(c), the time-averaged pressure distributions for Mach numbers of 0.71 and 0.905 exhibit a satisfactory correlation with the CFD data. Figure \ref{fig:cp_mean_and_rms}(d) shows the pressure fluctuations, revealing that both cases exhibit weaker unsteadiness compared to others. At Mach 0.71, vortex shedding dominates within the shock-free flow regime, while at Mach 0.905, stabilized shock waves prevail. In both cases, the neural network prediction exhibits a high degree of accuracy, closely aligning with the CFD data.

\begin{figure}
\centering
\includegraphics[width=0.75\linewidth]{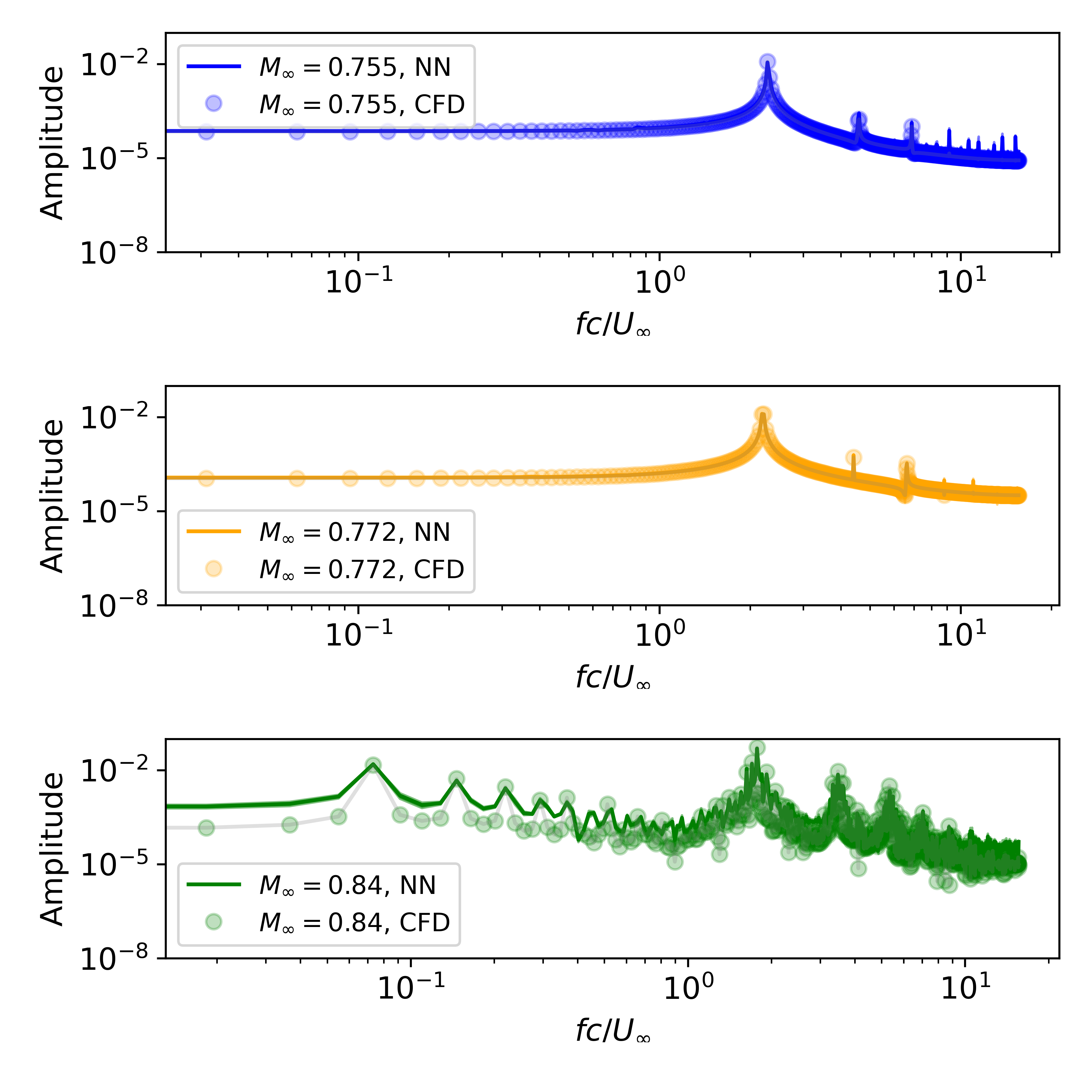}
\caption{Spectrum of the pressure signals using FFT. From top to bottom: $M_{\infty}=0.755$, $M_{\infty}=0.772$ and $M_{\infty}=0.84$. Here, ``NN'' represents the neural network, while ``CFD'' denotes the ground truth, i.e. CFD data.}
\label{fig:fft_intpl_pressure_probe}
\end{figure}

Figure \ref{fig:fft_intpl_pressure_probe} shows the FFT spectral of the pressure signals at the probe point ($x/c=0.972$, $y/c=-0.00452$). Each is evaluated using five models, and the shad shows $\pm 0.5\sigma$. The FFT spectral profiles reveal prominent frequencies occurring at approximately $fc/U_{\infty}=2.30$ for Mach number 0.755, $fc/U_{\infty}=2.20$ for Mach number 0.772, and $fc/U_{\infty}=1.77$ for Mach number 0.84. These frequencies are indicative of vortex shedding in the wake. Harmonics of these dominant frequencies manifest in the high-frequency spectrum. Notably, at Mach number 0.84, a shock wave is observed, oscillating alternately on the upper and lower aerofoil surfaces at a lower frequency. This behavior aligns with the occurrence of the peak value $fc/U_{\infty}=0.0733$ and its harmonics in the FFT profile.

\begin{figure}
\centering
\includegraphics[width=0.75\linewidth]{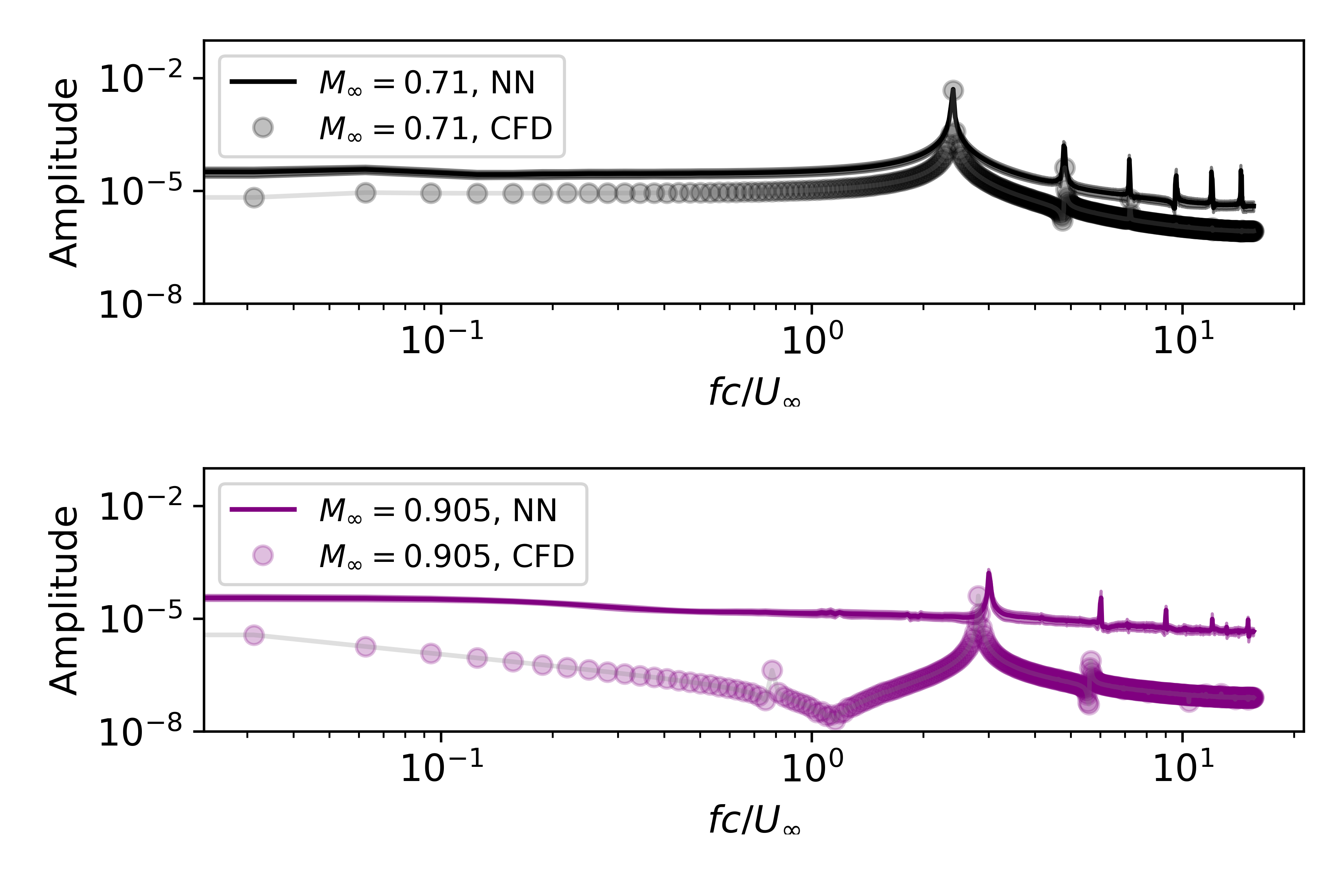}
\caption{Spectrum of the pressure signals using FFT. From top to bottom: $M_{\infty}=0.71$ and $M_{\infty}=0.905$. Here, ``NN'' represents the neural network, while ``CFD'' denotes the ground truth, i.e. CFD data.}
\label{fig:fft_extra_pressure_probe}
\end{figure}

Figure \ref{fig:fft_extra_pressure_probe} shows the FFT spectral of pressure signals for the extrapolation results. The test at Mach number 0.71 appears to be quite successful in terms of FFT spectrum predictions. The primary frequency and its harmonics closely resemble those observed in other cases without shock waves. The flow regime at Mach number 0.905, as mentioned previously, exhibits strong and stable shock waves, and the primary frequency values in the FFT spectrum remain very close to those in the CFD data. In conclusion, we can see that the neural network effectively replicates the frequency characteristics of the physical process.

%%%%% Inference %%%%%

\subsection{Inference from a time-averaged flowfield}
Moreover, we undertake a more demanding test for the trained model – predicting the evolution from the time-averaged flowfield. 
%While this scenario is infrequently encountered in computer graphics and computer vision applications,
While this scenario serves as an exceptionally tough and rigorous 'out-of-distribution' test, seldom encountered in other fields, its significance is pronounced in the realm of fluid dynamics.
%it holds significant importance in the field of fluid dynamics.

\begin{figure}
    \centering

\begin{subfigure}{.9\textwidth}
\centering
\includegraphics[width=\linewidth]{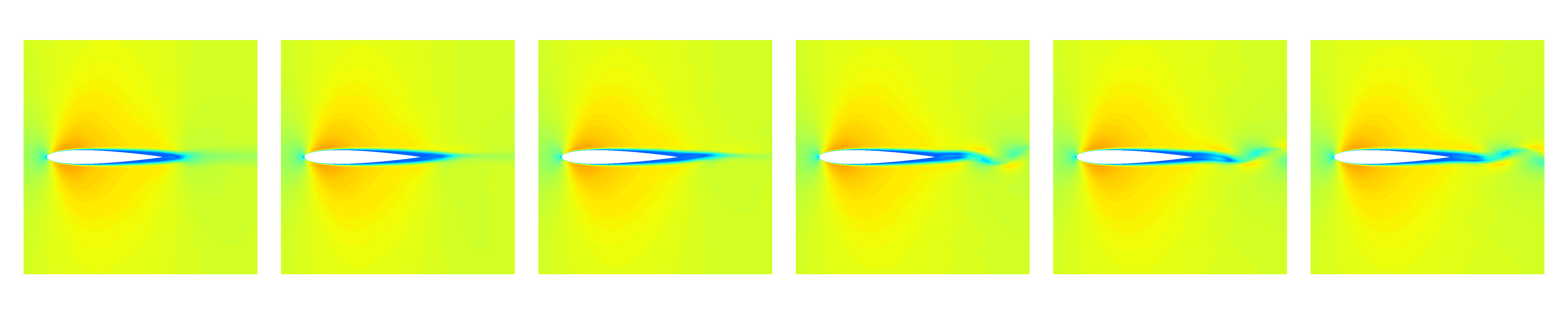}
\caption{$M_{\infty}=0.755$}
\end{subfigure}

\begin{subfigure}{.9\textwidth}
\centering
\includegraphics[width=\linewidth]{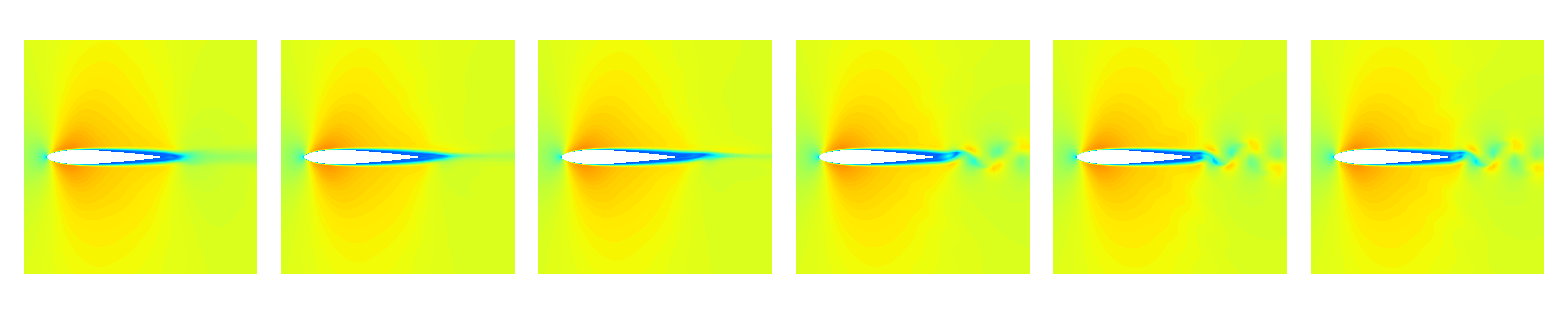}
\caption{$M_{\infty}=0.772$}
\end{subfigure}

\begin{subfigure}{.9\textwidth}
\centering
\includegraphics[width=\linewidth]{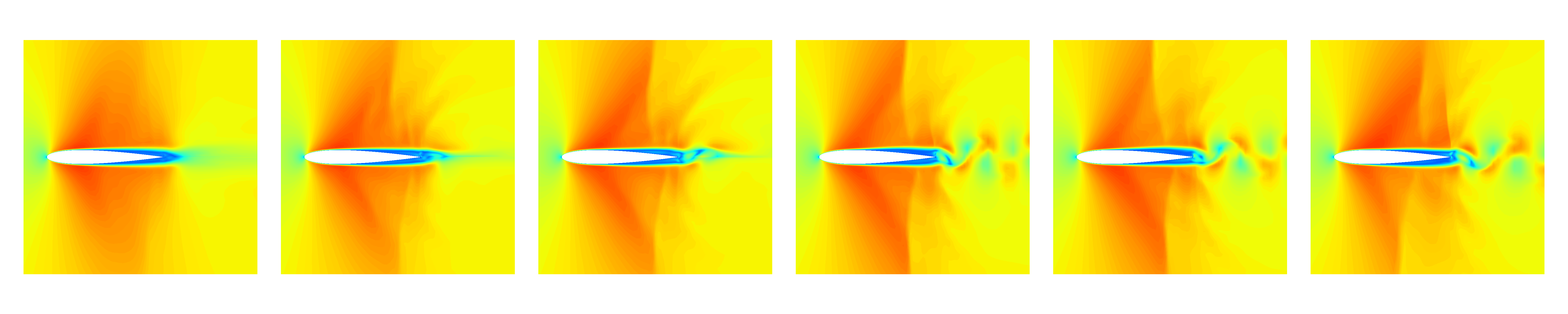}
\caption{$M_{\infty}=0.84$}
\end{subfigure}

\caption{The predicted evolution of the flow initialized using time-mean data at $M_{\infty}=0.755$, $M_{\infty}=0.772$ and $M_{\infty}=0.84$ (interpolation tests). From left to right, the Mach number contour maps are shown after 1, 6, 11, 51, 101, and 301 prediction steps.}
\label{fig_instability:evolution_from_mean_intpl}
\end{figure}

\begin{figure}
    \centering

\begin{subfigure}{.9\textwidth}
\centering
\includegraphics[width=\linewidth]{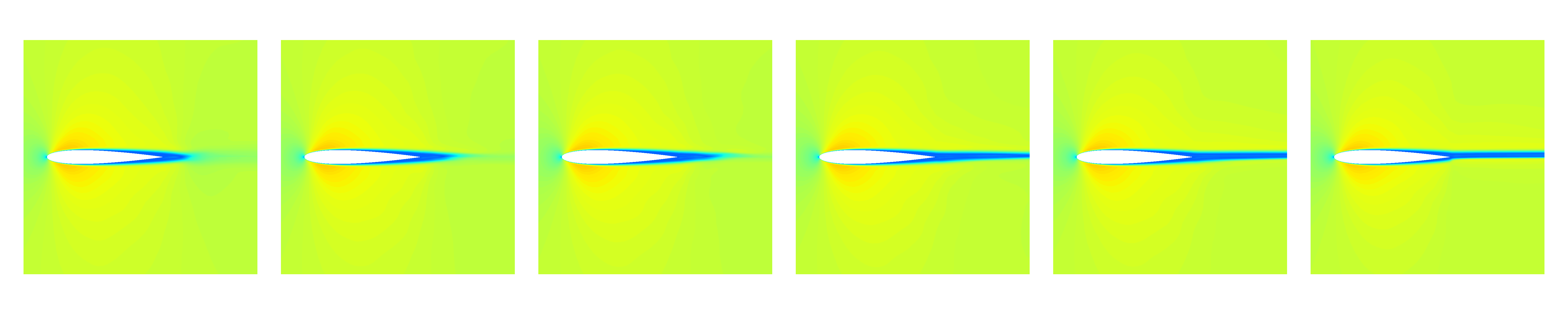}
\caption{$M_{\infty}=0.71$}
\end{subfigure}

\begin{subfigure}{.9\textwidth}
\centering
\includegraphics[width=\linewidth]{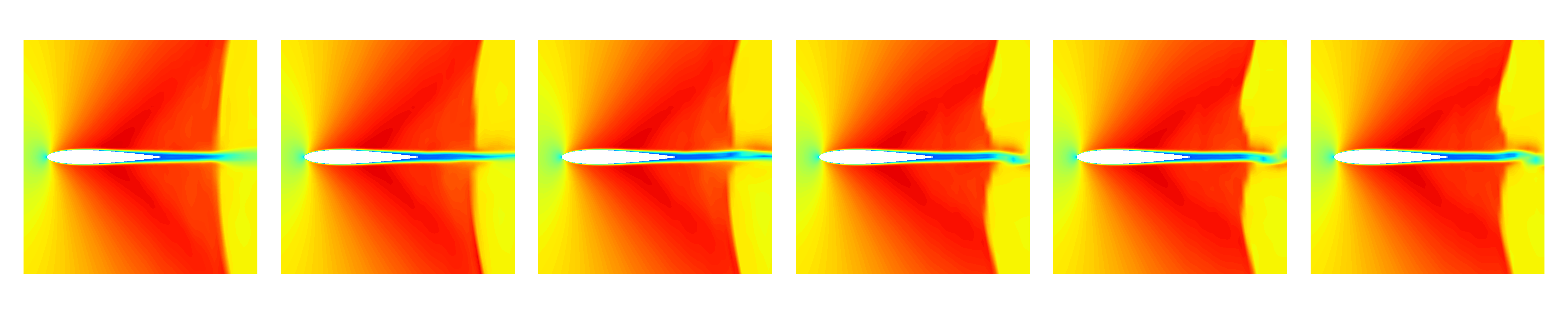}
\caption{$M_{\infty}=0.905$}
\end{subfigure}

\caption{The predicted evolution of the flow was initialized using time-mean data at $M_{\infty}=0.71$ and $M_{\infty}=0.905$ (extrapolation tests). From left to right, the Mach number contour maps are shown after 1, 6, 11, 51, 101, and 301 prediction steps.}
\label{fig_instability:evolution_from_mean_extra}
\end{figure}

We conduct inference using the learned model based on the time-averaged flow field for 5000 prediction steps, equivalent to 20000 simulation steps in the CFD simulations. Figure \ref{fig_instability:evolution_from_mean_intpl} illustrates three interpolation scenarios at Mach numbers 0.755, 0.772, and 0.84. In these scenarios, the learned model predicts the complete evolution of the flow field, showcasing the gradual development of flow instability from the initial time-averaged flow field to a fully developed, periodic unsteady flow field.
Moreover, Figure \ref{fig_instability:evolution_from_mean_extra} portrays two extrapolation scenarios corresponding to Mach numbers 0.71 and 0.905. These extrapolation cases pose significant challenges to the current methodology. Specifically, the model struggles to capture the unsteady evolution of the flow in the case of 0.71. In the case of 0.905, while the level of unsteadiness is minimal, the fully developed flow field structure remains acceptable, providing a promising outcome.

Having removed the transient stage data, we calculate the statistics, including the time-averaged pressure coefficient and fluctuations. Similar to the previous analysis, the FFT spectra of the pressure signals are measured at $x/c=0.972$, $y/c=-0.00452$, close to the aerofoil trailing edge.

\begin{figure}
    \centering
    
\begin{subfigure}{.45\textwidth}
\centering
\includegraphics[width=\linewidth]{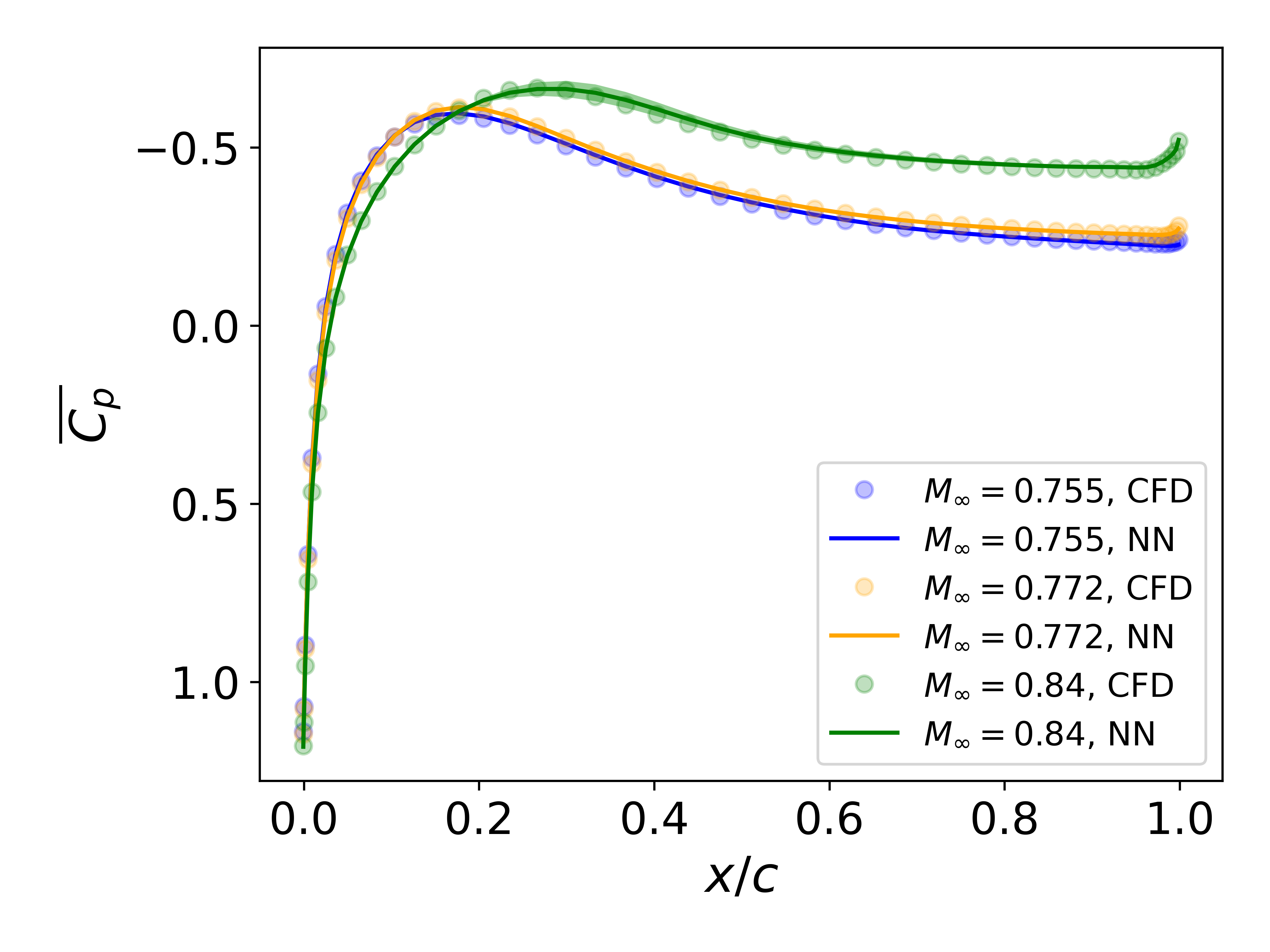}
\caption{}
\end{subfigure}
\begin{subfigure}{.45\textwidth}
\centering
\includegraphics[width=\linewidth]{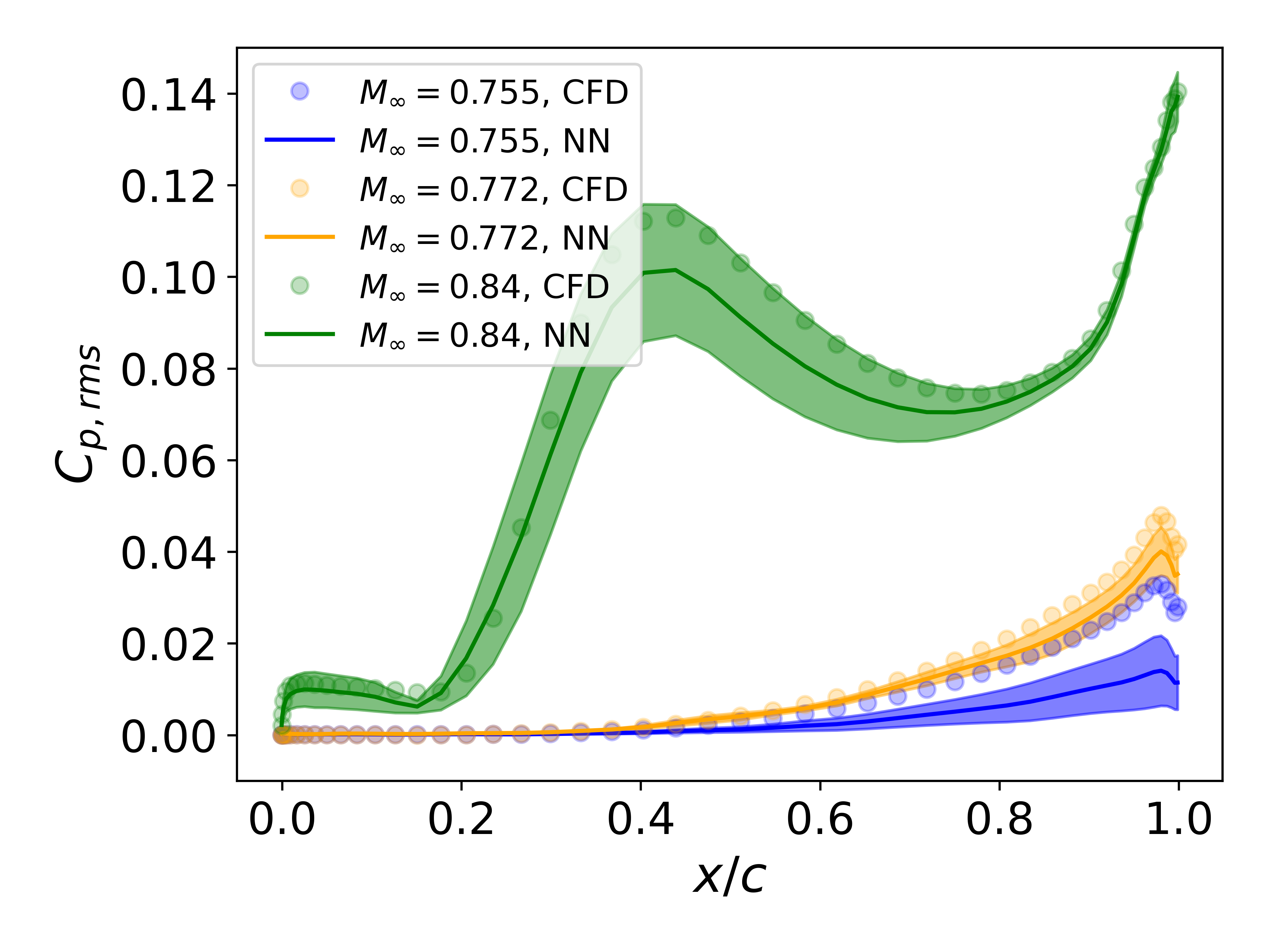}
\caption{}
\end{subfigure}

\begin{subfigure}{.45\textwidth}
\centering
\includegraphics[width=\linewidth]{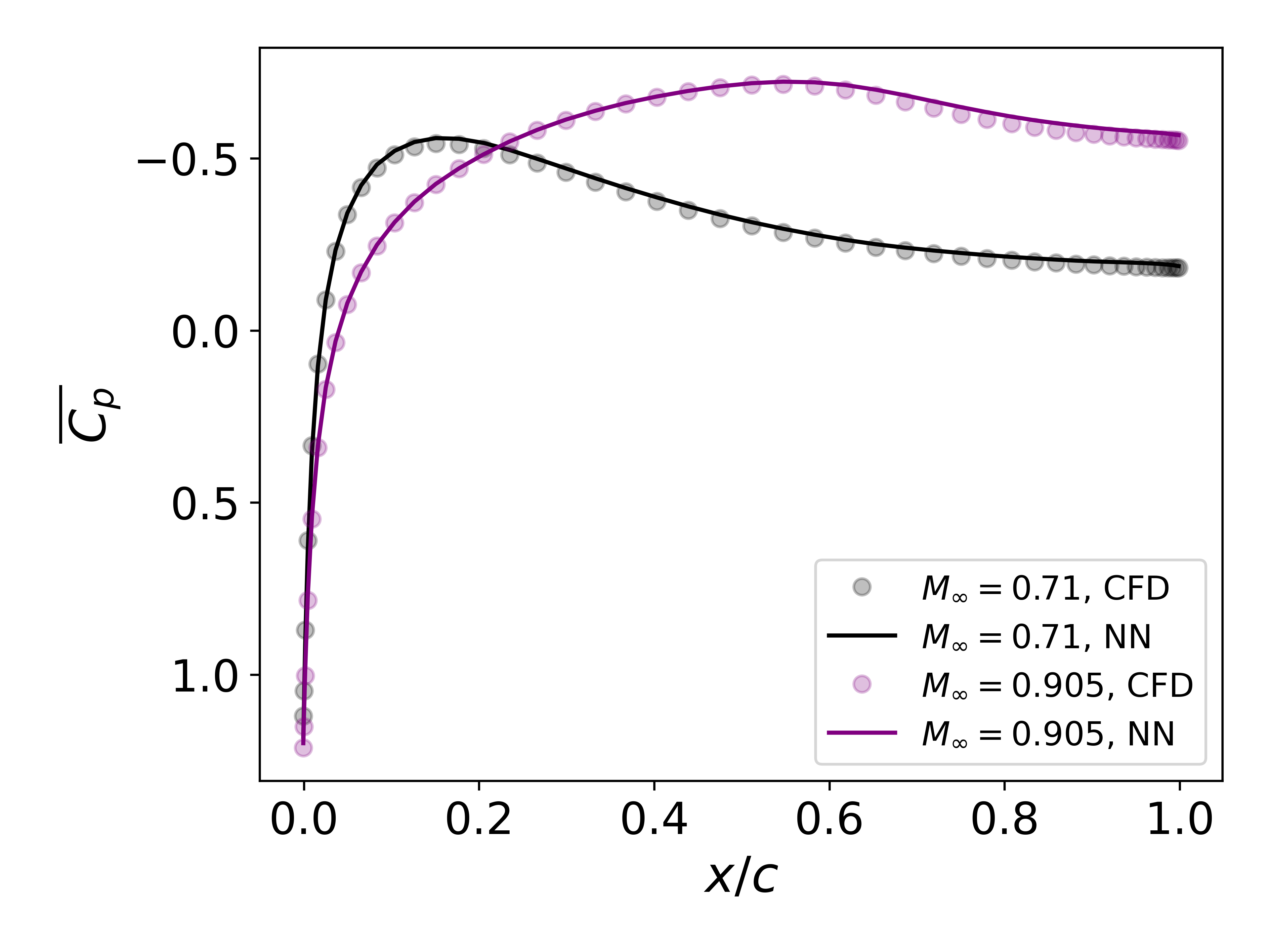}
\caption{}
\end{subfigure}
\begin{subfigure}{.45\textwidth}
\centering
\includegraphics[width=\linewidth]{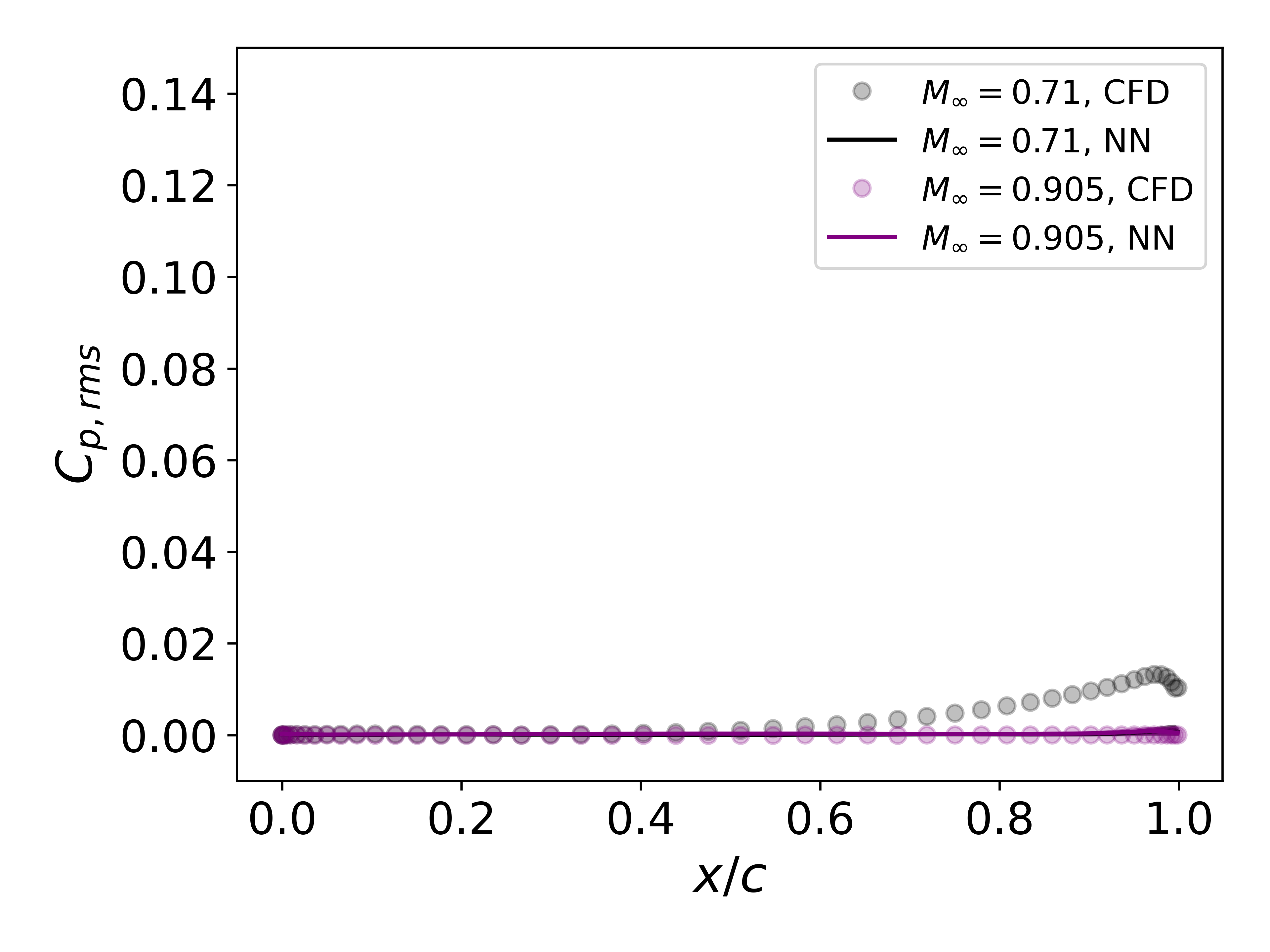}
\caption{}
\end{subfigure}

\caption{Pressure coefficients and fluctuations: (a-b) represent results from interpolation tests, while (c-d) depict outcomes from extrapolation tests. The inference is initialized with the time-mean data.}
\label{fig_instability:cp_mean_and_rms}
\end{figure}

Figure \ref{fig_instability:cp_mean_and_rms} presents the distribution of average surface pressure coefficients and pressure fluctuation, with $\pm0.5\sigma$ indicating the uncertainty intervals obtained from five models. From figure \ref{fig_instability:cp_mean_and_rms}(a), 
it is evident that 
%for the interpolation scenarios at Mach numbers 0.755, 0.772, and 0.84, 
the distribution of average pressure coefficients closely aligns with the CFD data, demonstrating a close and highly robust agreement (narrow uncertainty regions), although the pressure fluctuation distribution exhibits larger areas of uncertainty as shown in figure \ref{fig_instability:cp_mean_and_rms}(b).

The inference from the time-mean at extrapolation scenarios presents greater complexity. However, there are notable successes. As seen in Figure \ref{fig_instability:cp_mean_and_rms}(c), the distribution of average pressure coefficients exhibits robust performance for both Mach numbers.
Challenges arise at Mach number 0.71, as it appears that the pressure fluctuations on the wall have not 
been adequately stimulated, 
as shown in Figure \ref{fig_instability:cp_mean_and_rms}(d). 

%the overall trend in the average values remains promising, indicating the model's potential for effective extrapolation.

\begin{figure}
\centering
\includegraphics[width=0.85\linewidth]{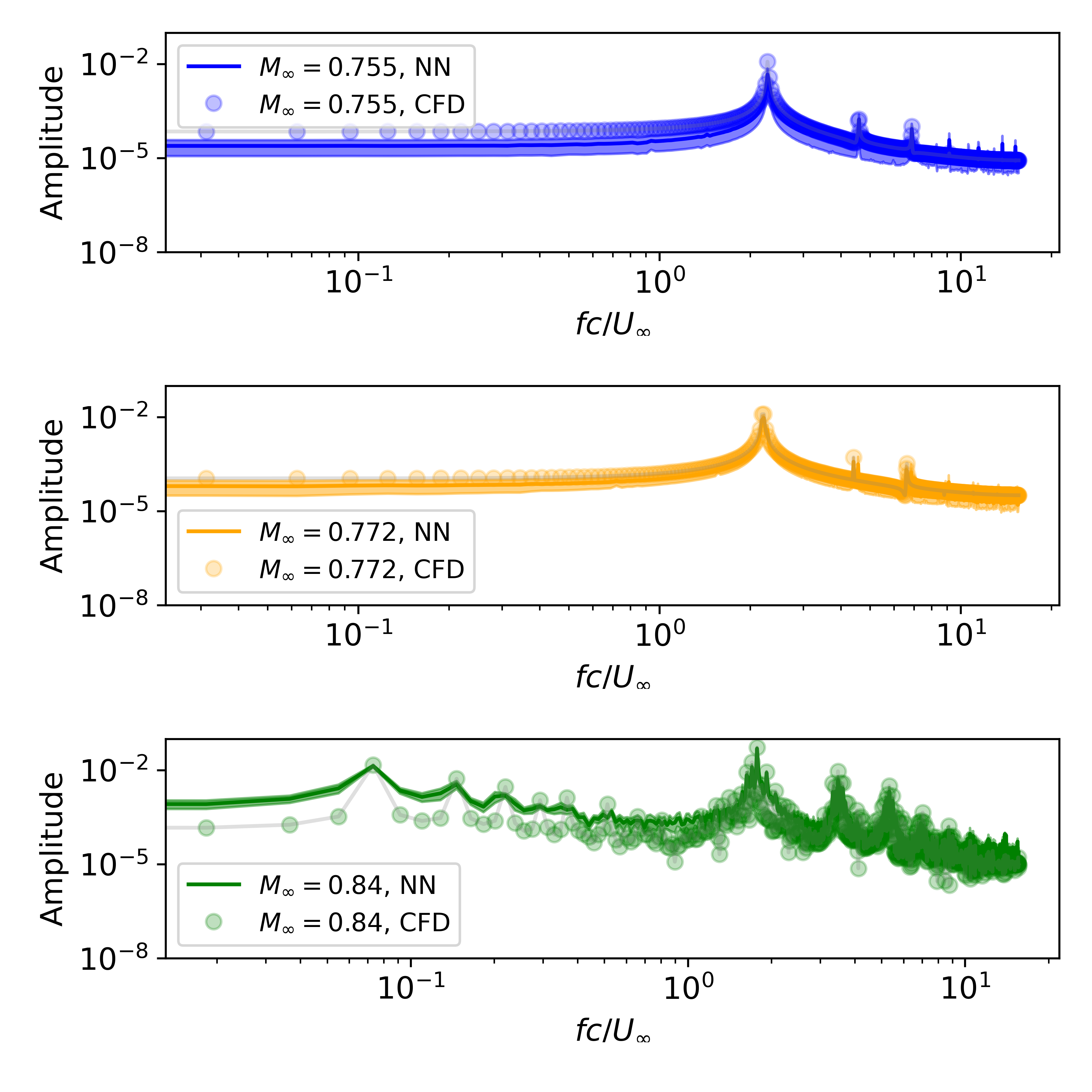}
\caption{Spectrum of the pressure signals using FFT. From top to bottom: $M_{\infty}=0.755$, $M_{\infty}=0.772$ and $M_{\infty}=0.84$. The inference is initialized using the time-mean data. Here, ``NN'' represents the neural network, while ``CFD'' denotes the ground truth, i.e. CFD data.}
\label{fig_instability:fft_intpl_pressure_probe}
\end{figure}

\begin{figure}
\centering
\includegraphics[width=0.85\linewidth]{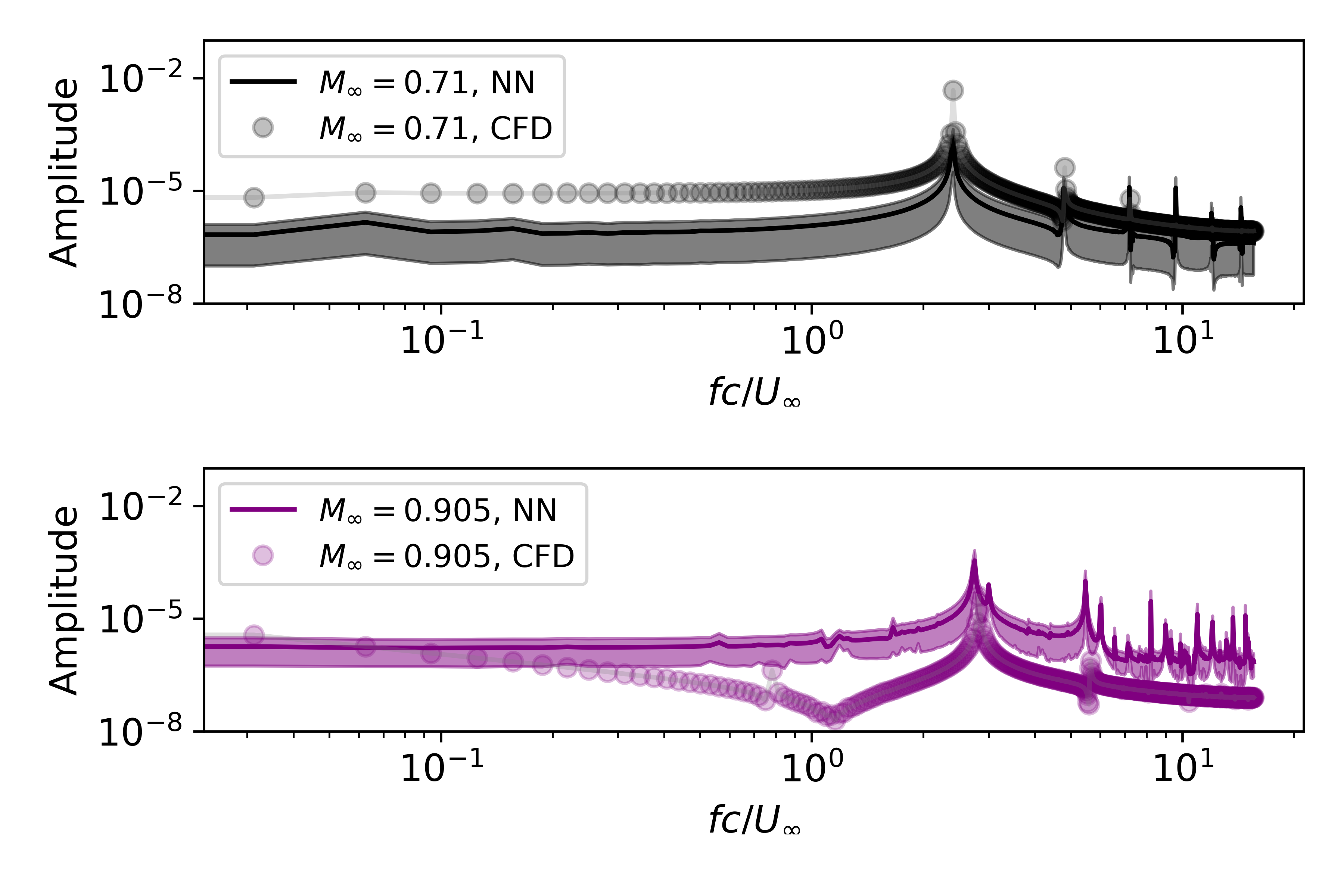}
\caption{Spectrum of the pressure signals using FFT. From top to bottom: $M_{\infty}=0.71$ and $M_{\infty}=0.905$. The inference is initialized using the time-mean data. Here, ``NN'' represents the neural network, while ``CFD'' denotes the ground truth, i.e. CFD data.}
\label{fig_instability:fft_extra_pressure_probe}
\end{figure}

As depicted in figure \ref{fig_instability:fft_intpl_pressure_probe}, the learned model demonstrates remarkable success in accurately capturing the peaks and harmonics in the FFT spectra at Mach numbers of 0.755, 0.772, and 0.84 when predicting from the time-averaged flow field. Although there is a slightly larger uncertainty region compared to the preceding tests, overall performance remains commendable. In the extrapolation experiments illustrated in figure \ref{fig_instability:fft_extra_pressure_probe}, while the FFT spectra distribution at Mach number 0.905 may exhibit some deviations from expectations, the model's performance remains relatively strong. Notably, at Mach number 0.71, where the absence of shock waves leads to high-frequency vortex shedding dominating low-frequency motion, the model initially appears to replicate this pattern accurately, akin to the previous tests. However, as observed earlier, the pressure fluctuations on the surface are 
not fully incited, % to prevent direct repetition of "adeq. stimulated"
%not adequately stimulated, 
resulting in an overall lower distribution of FFT signal compared to the CFD data.

In summary, we would like to re-emphasize that this test is exceptionally challenging, as it deviates significantly from conventional evaluations within the field of deep learning. Unlike typical tests that involve providing a transient flow field as an initial condition, this test begins with a time-averaged field and relies solely on a neural network's learned "flow evolution patterns" to predict subsequent developments and instabilities. While the model's performance appears suboptimal in extrapolation scenarios, it demonstrates excellent performance in interpolation cases. In these instances, the model accurately infers complete unsteady evolutions and successfully reproduces statistical quantities such as average pressure distribution and pressure fluctuation values.

%%%%%%%%%%%%%%%%%
%   post from mean 
%%%%%%%%%%%%%%%%%

%%%%% Inference %%%%%

\section{Training time and performance}
\noindent The performance of trained DNN models is one of the central factors motivating their use, in particular, in aerodynamic shape optimizations. We evaluate our models using a regular workstation with 12 cores, i.e. Intel\textsuperscript{\textregistered} Xeon\textsuperscript{\textregistered} CPU E5-1650 v4 @ 3.60GHz, with an NVidia GeForce GTX 1080 Ti GPU. Due to the difference in implementations, we compare solvers in terms of wall clock time, which is a pure computation time and is averaged over multiple runs without taking the start-up and initialization overheads into account.
A typical two-dimensional laminar simulation with CFL3D solver with the grid resolution $1024\times128$ requires 4000 time steps with a wall clock time of 21 hours using 8 CPU processes. 
Similarly, we assess the elapsed wall time required for 1000 predictions on a single batch of data, which equates to 4000 CFD steps, using neural network models of sizes 0.5M, 2.0M, and 8.1M. The respective wall times are approximately 7.01 seconds, 7.20 seconds, and 7.15 seconds. It is noteworthy that the runtime per solution can be notably reduced when evaluating multiple batch solutions simultaneously. Overall, The neural network inference has achieved considerable acceleration compared to CFD simulations.

\section{Learned-model-based global instability analysis}
\subsection{General description}
In the previous sections, it was noted that the learned model not only accurately forecasts the evolution of intricate flow fields under unseen conditions but also performs exceptionally well in more stringent tests, where the development of unsteady flow states stems from a time-averaged flow field. Encouraged by these test outcomes, we are motivated to explore a fundamental question: Can a model be trained such that it
%, through rigorous training, 
uncovers the underlying mechanisms of a dynamical system? If confirmed, the trained model de facto becomes a differentiable operator, encapsulating the dynamic characteristics present in the training dataset. To explore this question, this section utilizes the trained model as an operator for classical global stability analysis.

The choice of the base flow is crucial to any comprehensive analysis of global stability. The base flow serves as the reference point for linearizing the dynamics. However, the flow field is predominantly governed by instabilities for laminar and scale-resolving simulations, lacking a true stable flow pattern suitable for linearization. Nonetheless, as the vortex shedding and self-sustained shockwave (when buffet occurs) are periodic, we hence assume that the flow has statistically stationary properties over a long-time average. 
Note that this long-time-averaged flowfield is neither a solution of the governing equations nor a stationary point of the learned recursive operator,
The following global linear stability analysis makes sense in identifying the frequency content of large coherent structures 
rather than addressing the stability of the base flow \cite[]{Taira2017reviewModalAnalysis}.
%We should bear in mind that this study is based on a learned model and calculates the global mode for the time-mean base flow. 
%and there is no doubt that it will naturally evolve into an unsteady state. 
%Therefore, the present analysis does not focus on the stability of the base state, but rather on capturing large-scale coherent structures associated with specific frequencies.
\subsection{Discrete and continuous dynamical systems}
Because our neural network model is a discrete dynamical system (as in equation \ref{eqn:nn}), and stability analysis is typically based on continuous dynamical systems, we first need to establish a connection between discrete and continuous dynamical systems. This will enable us to then solve the corresponding eigenvalue problem. 

Firstly, we revisit the neural network equation \ref{eqn:nn} and transform it into \ref{eqn:nn_perturbation}, introducing the notation $\mathcal{N}$ under the assumption that $\theta$ and $M_{\infty}$ are predetermined:
\begin{equation*}\label{eqn:nn_revisit}
    f_{\theta}(\mathbf{q}_{t}, M_{\infty}):=\mathcal{N}(\mathbf{q}_t)
\end{equation*}
To clarify the discussion, we assume the time step $\Delta t$ is uniform and denote $\mathbf{q}_{t}$ as $\mathbf{q}_{n}$ and $\mathbf{q}_{t+\Delta t}$ as $\mathbf{q}_{n+1}$. Then we have
% some discussion could follow Li He s work
\begin{align}\label{eqn:nn_perturbation}
%        \mathbf{q}_{n+1}&=f_{\theta}(\mathbf{q}_n, M_{\infty}) \\
        \mathbf{q}_{n+1}&=\mathcal{N}(\mathbf{q}_n) \\
    \bar{\mathbf{q}}+ \mathbf{q}'_{n+1}&=\mathcal{N}(
\bar{\mathbf{q}}+ \mathbf{q}'_{n}) \\
\bar{\mathbf{q}}+ \mathbf{q}'_{n+1}&=\mathcal{N}(
\bar{\mathbf{q}})+ \frac{\partial\mathcal{N}}{\partial \mathbf{q}}\Bigg|_{\bar{\mathbf{q}}} \mathbf{q}'_{n}
\end{align}
Here, $\bar{\mathbf{q}}$ represents the time-averaged flow variables, calculated by 
\begin{align*}
          \bar{ \mathbf{q} }&=
\sum_{i=0}^{\infty}\mathbf{q}_{i+1}\\
&=\sum_{i=0}^{\infty}{\mathcal{N}(\mathbf{q}_i)}\\
&=\overline{\mathcal{N}(\mathbf{q}_i)} = \mathcal{N}(\bar{\mathbf{q}})-\mathbf{h} 
\end{align*}
where $\mathbf{h}$ represents the forcing term due to the nonlinearity of the operator. Hence, we have
\begin{align}
\mathbf{q}'_{n+1}&=\frac{\partial\mathcal{N}}{\partial \mathbf{q}}\Bigg|_{\bar{\mathbf{q}}} \mathbf{q}'_{n} + \mathbf{f} = \mathbf{L}(\bar{\mathbf{q}})\ \mathbf{q}'_{n} + \mathbf{h} \label{eqn:qtilde_update_neural} \\
\mathbf{L}(\bar{\mathbf{q}})&= \mathbf{V_L}\mathbf{\Sigma}\mathbf{V_L}^{-1} \label{eqn:eigenvectors_L_neural}
\end{align}
where $\mathbf{L}$ represents the Jacobian matrix of the operator $\mathcal{N}$, $\mathbf{V_L}$ is the eigenvector of $\mathbf{L}$, and $\mathbf{\Sigma}$ the eigenvalue.

On the other hand, looking at the continuous dynamical system (i.e. temporally continuous but discretized in space) is expressed as
\begin{align}
    \frac{d\mathbf{q}}{dt}=\mathcal{S}(\mathbf{q})
\end{align}
where $\mathcal{S}$ represents the nonlinear operator, such as the Navier-Stokes equation. Assuming $\mathbf{q}=\bar{\mathbf{q}}+\mathbf{q}'$,
the linearized ordinary differential equation in a continuous form can be written as
\begin{align}
\frac{d\mathbf{q}'}{dt}&=\frac{\partial\mathcal{S}}{\partial \mathbf{q}}\Bigg|_{\bar{\mathbf{q}}}\mathbf{q}' + \mathcal{S}(\bar{\mathbf{q}})\\
&=\mathbf{A}\mathbf{q}' + \mathcal{S}(\bar{\mathbf{q}})
\end{align}
The general solution of the nonhomogeneous equation is the sum of the homogeneous solution and the particular solution, and the discretized form can be expressed as
\begin{align}
\mathbf{q}'_{n+1}&=\mathrm{exp}({\mathbf{A}\Delta t})\mathbf{q}'_{n} + \mathbf{f} \label{eqn:qtilde_update_ref}\\
&=\mathbf{V_A}\mathrm{exp}(\mathbf{\Lambda}\Delta t)\mathbf{V_A}^{-1} \mathbf{q}'_n + \mathbf{f}  \\
\mathbf{A}&=\mathbf{V_A}\mathbf{\Lambda}\mathbf{V_A}^{-1} \label{eqn:eigenvectors_A_ref}
%\mathbf{L}(\bar{\mathbf{q}})&= \mathbf{V_L}\mathbf{\Sigma}\mathbf{V_L}^{-1}
\end{align}
Here, $\mathbf{A}$ represents the Jacobian matrix of the continuous dynamical system, $\mathbf{V_A}$ is the eigenvector of $\mathbf{A}$, and $\mathbf{\Lambda}$ the eigenvalue.
%Looking at the 
The similarity of the two discrete dynamical system recurrence formulas, equations \ref{eqn:qtilde_update_neural} and \ref{eqn:qtilde_update_ref}, means that 
%if a learned model that was trained on high-fidelity data is sufficiently accurate, it can guarantees a high degree of validity with $\mathbf{L}=\mathbf{A}$. This equivalence holds on the premise that the data-driven model accurately captures the dynamics of the system.
If a learned model trained on high-fidelity data achieves sufficient accuracy, it can provide a high degree of validity, potentially leading to $\mathbf{L}=\mathbf{A}$ in the limit. This equivalence is contingent upon the data-driven model accurately capturing the dynamics of the system.

Hence, from equations \ref{eqn:eigenvectors_L_neural} and \ref{eqn:eigenvectors_A_ref}, we have 
\begin{align}
    \lambda_i = \mathrm{log}(\sigma_i)/\Delta t
\end{align}
where $\lambda_i$ are the eigenvalues of $\mathbf{A}$ and $\sigma_i$ the eigenvalues of $\mathbf{L}$. 
Under the assumption of a highly accurate data-driven model, 
the eigenvectors $\mathbf{V_L}$ can be regarded as well-estimated counterparts to $\mathbf{V_A}$. 
%
%This implies that without relying on the governing equations of a continuous dynamical system (e.g. equation \ref{eqn:qtilde_update_ref}), as long as we can train an accurate neural network using high-fidelity data, we can analyze the dynamics of the original continuous system by examining the dynamical characteristics of the trained model. The insights gained from the properties of $\mathbf{L}$ provide information about the original dynamical system.
This implies that, as long as we can train a sufficiently accurate neural network, we can analyze the dynamics of the original continuous system by only examining the dynamical characteristics of the trained model. This is possible without relying on the governing equations of a continuous dynamical system such as equation \ref{eqn:qtilde_update_ref}.
Instead, insights about the original dynamical system can be gained by analyzing $\mathbf{L}$.

The eigenvalues of $\mathbf{A}$ are associated with the system's natural frequencies. The real parts of the eigenvalues indicate the exponential growth or decay rates, while the imaginary parts correspond to oscillatory behavior. Nonetheless, learned models usually come with uncertainty, which is typically caused by random initialization of parameters and mini-batch shuffling. To obtain a reliable Jacobian matrix $\mathbf{A}$, we employ ``model ensembling'' by combining five trained models which are initialized with different random seeds, and calculate the ensemble average $\mathbf{A}$. The ensemble approach captures a broader range of multiple solutions, 
%resulting in a more stable and representative average gradient, 
and reduces sensitivity to specific weight initializations, 
resulting in a more stable and representative average gradient. 
This approach has been widely used in applications such as learned model-based control and inverse design tasks \cite[]{chua2018deep, allen2022inverse}.
%mproving generalization and overall reliability in gradient computations. 

\subsection{Eigenvalue spectrum and dominant modes}
\begin{figure}
\centering

\tabskip=0pt
\valign{#\cr
  \hbox{%
    \begin{subfigure}[b]{.38\textwidth}
    \centering
    \includegraphics[width=\textwidth]{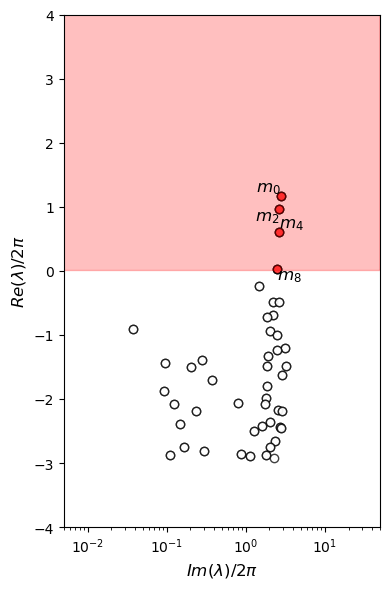}
    \caption{Eigenvalue spectrum}
    \end{subfigure}%
  }\cr
  \noalign{\hfill}
  \hbox{%
    \begin{subfigure}{.6\textwidth}
    \centering
    \includegraphics[height=3.5cm,width=\textwidth]{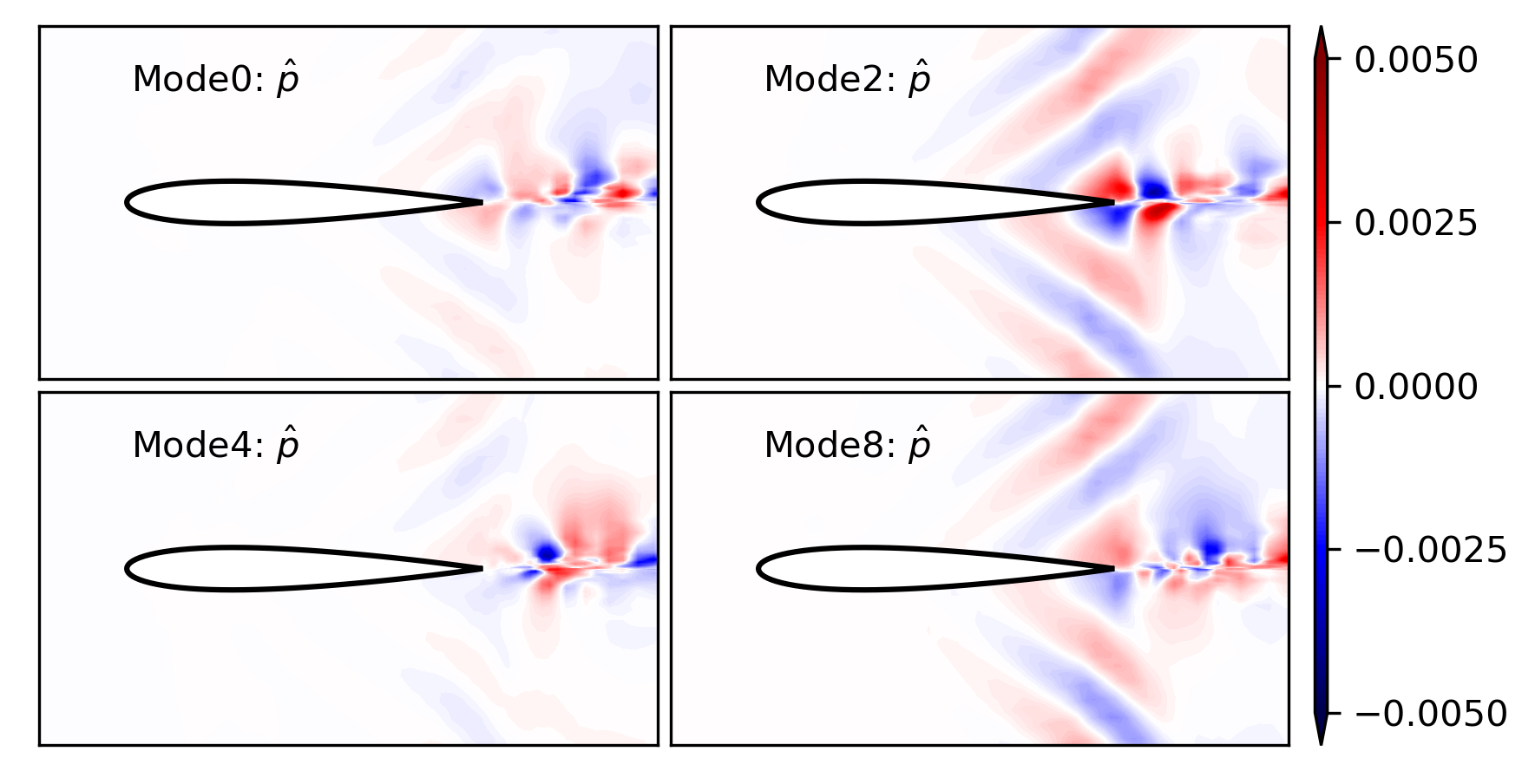}
    \caption{Pressure modes, real part.}
    \end{subfigure}%
  }\vfill
  \hbox{%
    \begin{subfigure}{.6\textwidth}
    \centering
    \includegraphics[height=3.5cm,width=\textwidth]{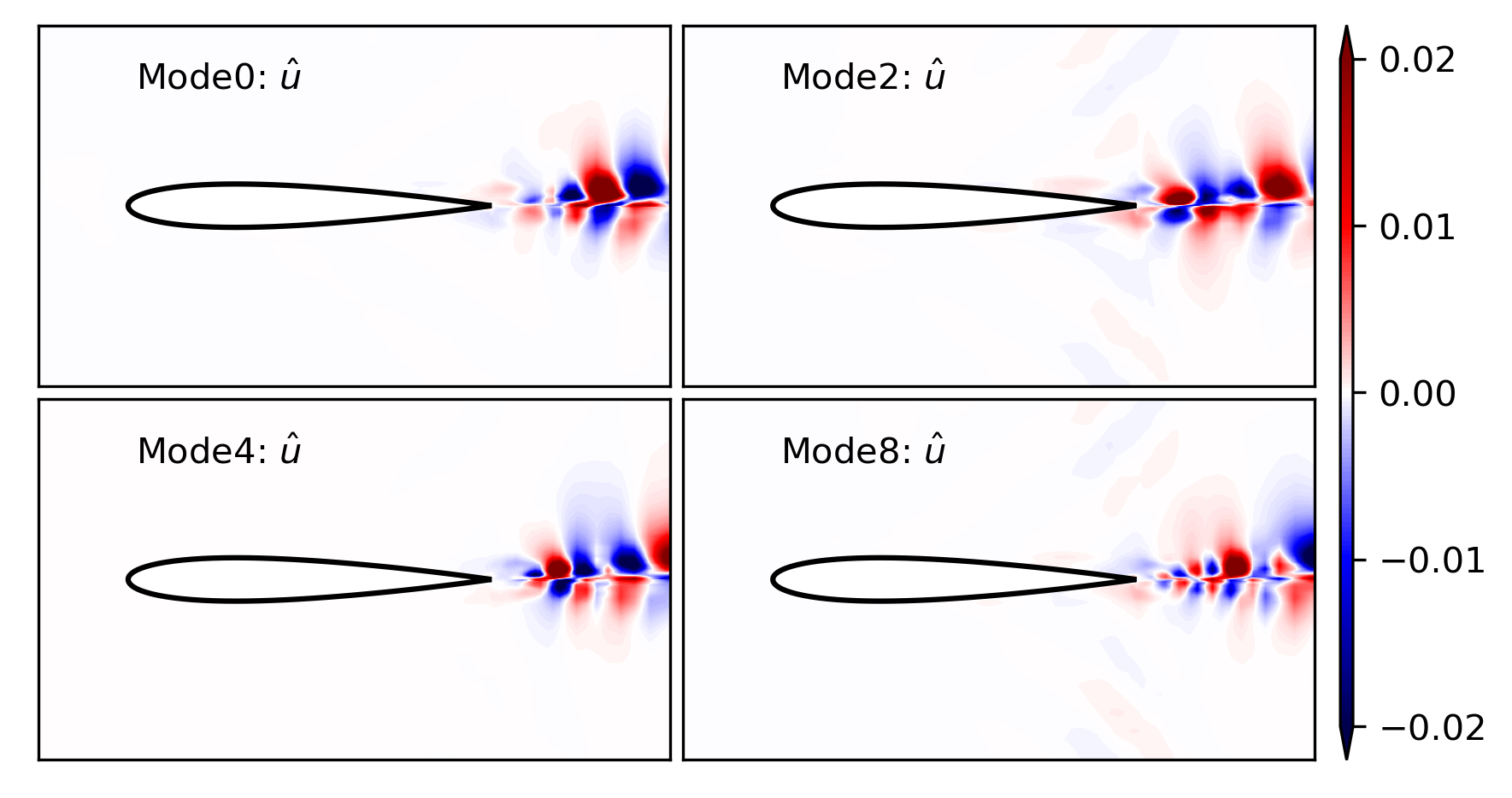}
    \caption{Stream-wise modes, real part}
    \end{subfigure}%
  }\cr
}

\caption{$M_{\infty}=0.772$}
\label{fig:eigen_plane_0p772}
\end{figure}

%
%Understanding these frequencies is crucial in predicting the system's response to different inputs. 
Firstly, we look at the eigenvalue spectrum plot at $M_{\infty}=0.772$.
In figure \ref{fig:eigen_plane_0p772}(a), the eigenvalue spectrum exhibits a large cluster of points in the region with negative real parts, and several eigenvalues extend from the stable half plane to the unstable half plane, indicating that the system is in a critical stability state. In such cases, the system might undergo a transition from stability to instability or exhibit oscillatory behavior.

%Temporal frequencies of the modes, and the relative errors between the modes and the frequencies identified in the FFT (see Fig.XXX) are calculated. The dominant peak in the FFT corresponds to mode M1, with a relative frequency error of 1.82\%, while modes M0, M2, and M3 all have errors below 3\%. Note, however, that the most unstable mode, M0, does not correspond to the dominant peak in the FFT, suggesting that important nonlinear effects, that are not captured by a modal analysis, play a role in selecting the frequency.

Temporal frequencies of the modes, and the relative errors between the modes and the frequencies identified in the FFT (see figure \ref{fig:fft_intpl_pressure_probe}) are given by table \ref{tab:eigenvalues_mach_0p772} The dominant peak in the FFT corresponds to unstable modes $m_0$ to $m_8$, with a relative frequency error of above 12\%, suggesting that significant nonlinear effects play a role in selecting the frequency.
%We pay attention to the point at $m=7,8$, where the real part of the eigenvalue is near zero and the imaginary part $\pm\mathrm{Im}\{\frac{\lambda}{2\pi}\}=2.48$, indicating the system exhibits oscillatory behavior at $2.48$, which is close to the high-frequency peak $2.2$ measured in the flowfield.
\begin{table}
    \centering
    \begin{tabular}{l c c c }
        %\toprule
        \hline
Mode & Eigenvalue $\frac{\lambda}{2\pi}$ & FFT frequency & $\Delta \mathrm{Im}\{\frac{\lambda}{2\pi}\}$\\
        \hline
$m_{0}$ & $1.1703308+2.81216j$ & 2.20& 27.82\%\\
$m_{2}$ & $0.9641248+2.6544704j$ & 2.20& 20.66\%\\
$m_{4}$ & $0.6119804+2.647502j$ &  2.20 &  20.3\% \\
$m_{8}$ & $0.034572423+2.483544j$ & 2.20& 12.89\%

 \end{tabular}
\caption{Selected frequencies for $M_{\infty}=0.772$}
 \label{tab:eigenvalues_mach_0p772}
\end{table}

\iffalse
Lambda/(2*pi) 0 (2.8078394+0.27758664j)
Lambda/(2*pi) 1 (2.8078394-0.27758664j)
Lambda/(2*pi) 2 (2.274273+0.15518315j)
Lambda/(2*pi) 3 (2.274273-0.15518315j)
Lambda/(2*pi) 4 (2.2582438+0.29768786j)
Lambda/(2*pi) 5 (2.2582438-0.29768786j)
Lambda/(2*pi) 6 (2.0605779+2.3777623j)
\fi

\begin{table}
    \centering
    \begin{tabular}{l c c c }
        %\toprule
        \hline
Mode & Eigenvalue $\frac{\lambda}{2\pi}$ & FFT frequency & $\Delta \mathrm{Im}\{\frac{\lambda}{2\pi}\}$\\
        \hline
$m_{12}$ & $2.8078394+0.27758664j$ & \/&  \\
$m_{70}$ & $-0.06118219+\mathbf{0.09932874}j$ & 0.077 & 28.57\% \\
$m_{72}$ & $-0.06684557+0.3259675j$ &  \/ &  \\
$m_{74}$ & $-0.07541487+\mathbf{1.9026724}j$ & 1.77 & 7.34\%

 \end{tabular}
 \caption{Selected frequencies for $M_{\infty}=0.84$}
 \label{tab:eigenvalues_mach_0p84}
\end{table}

\iffalse
Lambda/(2*pi) 70 (-0.06118219+0.09932874j)
Lambda/(2*pi) 71 (-0.06118219-0.09932874j)
Lambda/(2*pi) 72 (-0.06684557+0.3259675j)
Lambda/(2*pi) 73 (-0.06684557-0.3259675j)
Lambda/(2*pi) 74 (-0.07541487+1.9026724j)
Lambda/(2*pi) 75 (-0.07541487-1.9026724j)
Lambda/(2*pi) 76 (-0.12535489-3.4778643j)
\fi

In figure \ref{fig:eigen_plane_0p772}(b), the pressure modal patterns of modes $m_0$ to $m_8$ exhibit a clear alternation between positive and negative values, indicating the presence of pressure wave propagation. Particularly noteworthy is the discernible propagation from the airfoil trailing edge towards the far-field direction. Notably, within the wake region downstream, the pattern arrangement suggests oscillations in pressure associated with vortex shedding.
This interpretation finds support in the stream-wise velocity modes (i.e. the real part of $\hat{u}$) depicted in figure \ref{fig:eigen_plane_0p772}(c), where it becomes apparent that the dominant mode driving instability within the flow system corresponds to vortex shedding.

%In contrast, looking at the eigenvalue spectrum at $M_{\infty}=0.84$ in figure \ref{fig:eigen_plane_0p84}(a), two eigenvalues in the unstable half-plane correspond to the high-frequency Kelvin-Helmholtz instability, and some of these marginally stable modes in association with branchlike structures that are more heavily damped than the “root” mode located on $\mathbf{Re}\{\lambda\}=0$.

\begin{figure}
\centering

\tabskip=0pt
\valign{#\cr
  \hbox{%
    \begin{subfigure}[b]{.35\textwidth}
    \centering
    \includegraphics[width=\textwidth]{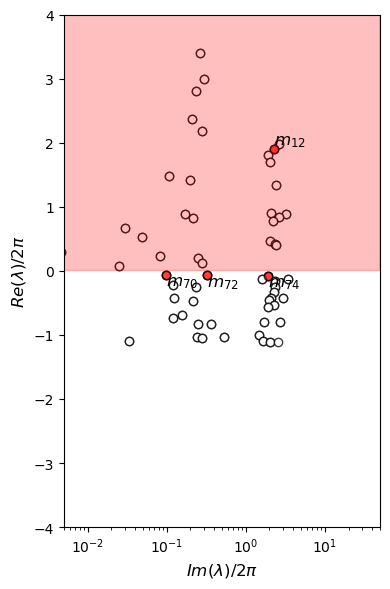}
    \caption{Eigenvalue spectrum}
    \end{subfigure}%
  }\cr
  \noalign{\hfill}
  \hbox{%
    \begin{subfigure}{.6\textwidth}
    \centering
    \includegraphics[height=3.5cm,width=\textwidth]{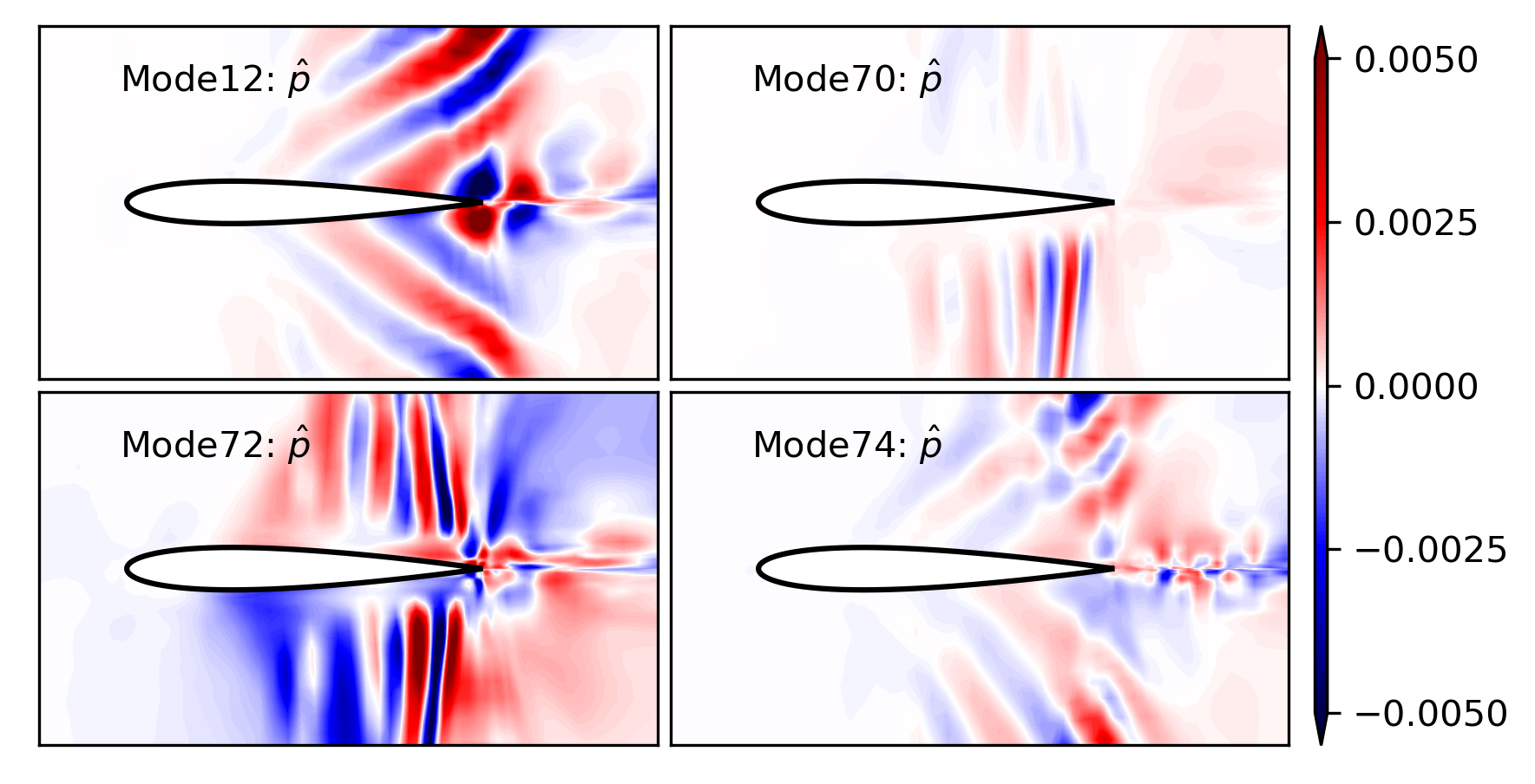}
    \caption{Pressure modes, real part.}
    \end{subfigure}%
  }\vfill
  \hbox{%
    \begin{subfigure}{.6\textwidth}
    \centering
    \includegraphics[height=3.5cm,width=\textwidth]{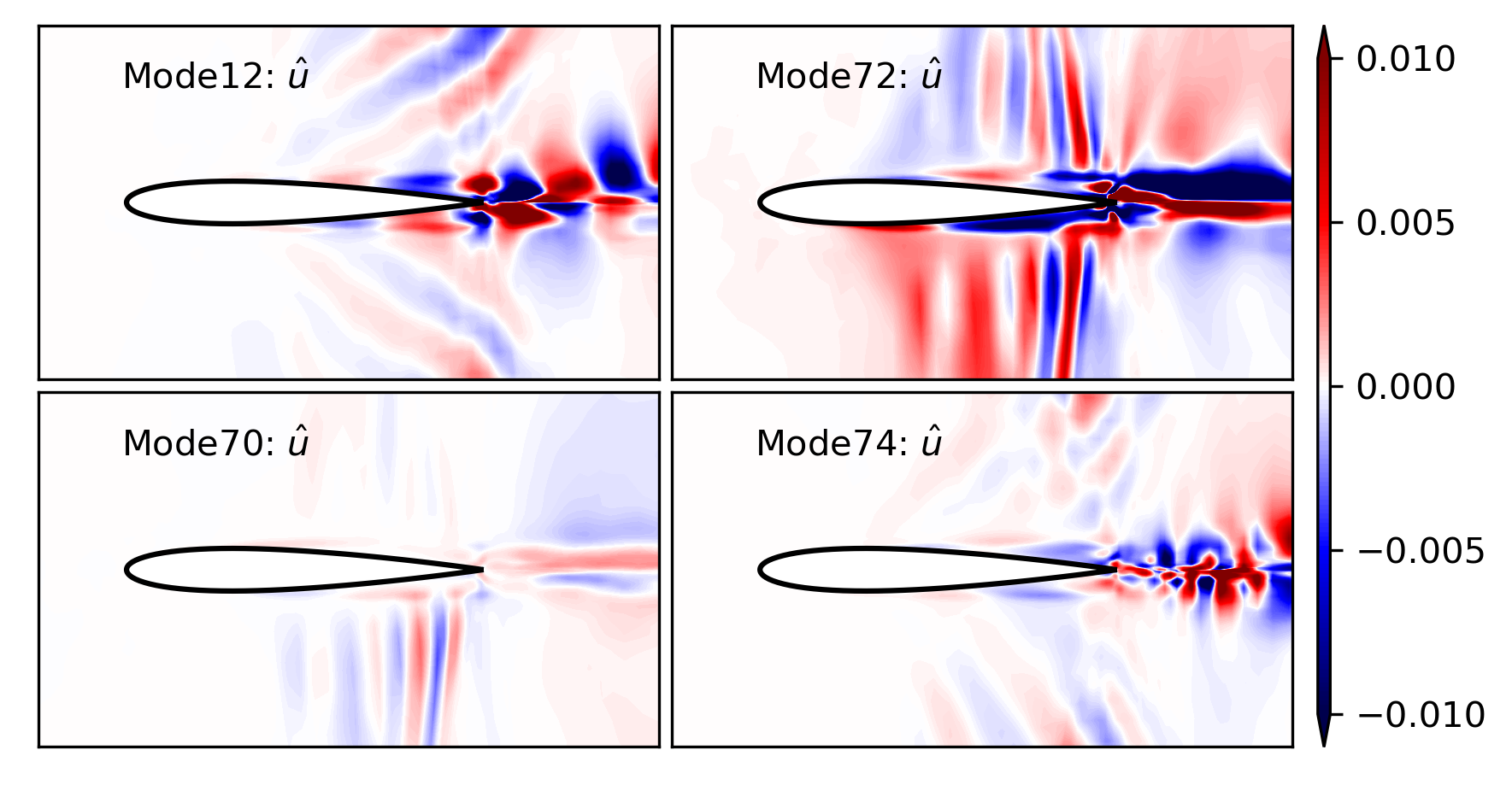}
    \caption{Stream-wise modes, real part}
    \end{subfigure}%
  }\cr
}

\caption{$M_{\infty}=0.84$}
\label{fig:eigen_plane_0p84}
\end{figure}

Upon examining the case of Mach number 0.84, it is observed that the eigenvalue spectrum exhibits multiple branches extending towards the unstable upper half-plane. We particularly focus on modes 12, 70, 72, and 74. As depicted in figure \ref{fig:eigen_plane_0p84}(a), the branches corresponding to modes 12 and 74 represent high-frequency signals (i.e., the imaginary parts of the eigenvalues). The associated pressure modal distributions in figure \ref{fig:eigen_plane_0p84}(b) and velocity modal distributions in figure \ref{fig:eigen_plane_0p84}(c) resemble those observed at Mach number 0.772, indicating the occurrence of vortex shedding resulting from Kelvin-Helmholtz instability. Furthermore, the alternating arrangement of positive and negative pressure modal patterns (see figure \ref{fig:eigen_plane_0p84}(b)) reveals the direction of pressure wave propagation from the airfoil trailing edge towards the far field.

Unlike Mach number 0.772, the case at Mach number 0.84 exhibits additional low-frequency unstable branches. To investigate this, we examined modes 70 and 72 at the axis $\mathbf{Re}{\lambda}=0$. The frequencies corresponding to these two modes are 0.077 and 1.77, respectively, likely associated with low-frequency peaks observed in FFT analysis of the flow field. As analyzed earlier, these frequencies correspond to the low-frequency motion of shock waves following buffet onset. The pressure modal distributions in figure \ref{fig:eigen_plane_0p84}(b) and velocity modal distributions in figure \ref{fig:eigen_plane_0p84}(c) for modes 70 and 72 reveal stripe structures parallel to the flow direction. In the trailing edge region, these modes correspond to large-scale spatial structures, consistent with the low-frequency characteristics of shock wave motion.

%{
%[because this is such an important section, it would be neat to summarize the results, i.e., have the goals from the beginning of the section been reached? along the lines of:]

%These results show that by treating the trained networks as differentiable operators, we can successfully analyze the dynamic characteristics characterizing the training dataset with global stability analysis. In contrast to classical approaches, this does not require reformulations of a continuous description, but instead is made possible by only relying on a dataset for training. At the same time it shows that global stability analysis is a suitable tool to interpret the learned dynamics of neural networks.
%...
%maybe also faster?
%...
%}

In summary, by treating the trained neural networks as differentiable operators, we conducted a global stability analysis to explore the complex dynamics embedded within the training dataset. This method diverges from traditional approaches, as it relies solely on training data rather than continuous system formulations. Our findings highlight the versatility of neural networks in capturing system behavior and demonstrate the applicability of global stability analysis in interpreting the learned dynamics of neural networks. This approach provides valuable insights into the underlying mechanisms of complex systems and contributes to the growing intersection of machine learning and classical stability analysis methodologies.

%In our pursuit of improved performance in calculating the Jacobian matrix, we combined five trained models, and each initialized with different random seeds, known as Ensembling neural networks. Due to factors like random initialization and optimization stochasticity, individual networks may produce diverse predictions. The ensemble approach captures a broader range of solutions by training multiple networks with distinct initializations, resulting in a more stable and representative average gradient. This strategy reduces sensitivity to specific weight initializations, improving generalization and overall reliability in gradient computations. Note that ensembles are widely preferred in applications such as learned model-based control and inverse design tasks 
%\cite[]{chua2018deep, allen2022inverse}, offering robust predictive quality.

%\iffalse

%\fi

%

\section{Concluding remarks}\label{sec:concluding_remarks}
In this study, 
we propose a novel approach for data-driven prediction of transonic flow fields. 
We trained attention U-Net models using roll-out training and noise addition techniques.
The learned models can predict the evolution of complex flows under conditions that were not presented during training, including initial instantaneous flow fields at unseen Mach numbers, and even with time-averaged flow fields. We demonstrate that the neural network model not only qualitatively predicts the subsequent evolution of the flow field but also quantitatively forecasts the aerodynamic characteristics of the airfoil, including average pressure distribution, fluctuating pressure distribution, and the frequency spectrum.

We suggest that a successful learned model should be able to replicate the characteristics of dynamical systems. 
We have found, through classic Jacobian matrix-based global stability analysis, that the learned model can obtain unstable eigenvalues and eigenvectors, 
and the mode shapes and frequencies were found to match very well with the
FFT analysis of the pressure signals. 
In the shock-free case at Mach number 0.772, the eigenvalue spectrum plot shows a high-frequency unstable branch, corresponding to the Kelvin-Helmholtz mode. In the buffet-onset case at Mach number 0.84, the eigenvalue spectrum exhibits multiple unstable branches, associated with the high-frequency Kelvin-Helmholtz mode and the low-frequency shock wave motion.

In general, our work contributes to the broader goal of enhancing the interpretability of neural network models, as it offers a systematic methodology for analyzing and interpreting the behavior of such models in the context of complex dynamical systems.
The integration of deep learning and global instability analysis stands as a promising methodology for addressing the complexities of flow control and shape optimization. The neural-network-based resolvent analysis will be an interesting direction.
%Furthermore, the application of neural networks in learning physical dynamical systems is closely intertwined with the concept of AI interpretability, thereby fostering the potential progression of deep learning techniques.

\section*{Code and data availability}
To ensure reproducibility, the full code, and training as well as test datasets for this work will be made available upon publication.

\section*{Acknowledgements} 
This work was supported by the European Research Council Consolidator Grant \emph{SpaTe}
(Grant agreement ID: 863850).

\section*{Declaration of interests} 
The authors report no conflict of interest.

\appendix
\renewcommand\thefigure{\thesection\arabic{figure}}

\renewcommand{\thetable}{\thesection\arabic{table}}

\section{}\label{appA}
\setcounter{figure}{0}    
\setcounter{table}{0}

\iffalse
\begin{table}
\centering
\begin{tabular}{c c c c }
\hline
Layer & Type & Details & Output Size \\
\hline
\hline
1 & Convolutional & Kernel Size: 3x3, Filters: 64 & (256x64, 64) \\
\hline
2 & Convolutional & Kernel Size: 3x3, Filters: 64 & (256x64, 64) \\
\hline
3 & Max-Pooling & Pool Size: 2x2 & (128x32, 64) \\
\hline
4 & Convolutional & Kernel Size: 3x3, Filters: 128 & (128x32, 128) \\
\hline
5 & Convolutional & Kernel Size: 3x3, Filters: 128 & (128x32, 128) \\
\hline
6 & Max-Pooling & Pool Size: 2x2 & (64x16, 128) \\
\hline
7 & Attention Gate & - & (64x16, 128) \\
\hline
8 & Up-sampling & Scale Factor: 2 & (128x32, 128) \\
\hline
9 & Concatenation & - & (128x32, 256) \\
\hline
10 & Convolutional & Kernel Size: 3x3, Filters: 64 & (128x32, 64) \\
\hline
11 & Convolutional & Kernel Size: 3x3, Filters: 64 & (128x32, 64) \\
\hline
12 & Attention Gate & - & (128x32, 64) \\
\hline
13 & Up-sampling & Scale Factor: 2 & (256x64, 64) \\
\hline
14 & Concatenation & - & (256x64, 128) \\
\hline
15 & Convolutional & Kernel Size: 3x3, Filters: 32 & (256x64, 32) \\
\hline
16 & Convolutional & Kernel Size: 3x3, Filters: 4 & (256x64, 4) \\
\hline
\end{tabular}
\caption{Neural Network Architecture}
\label{tab-App:network_architecture_details}
\end{table}
\fi

\begin{figure}
    \centering
\begin{subfigure}{1.\textwidth}
\centering
\includegraphics[width=\linewidth]{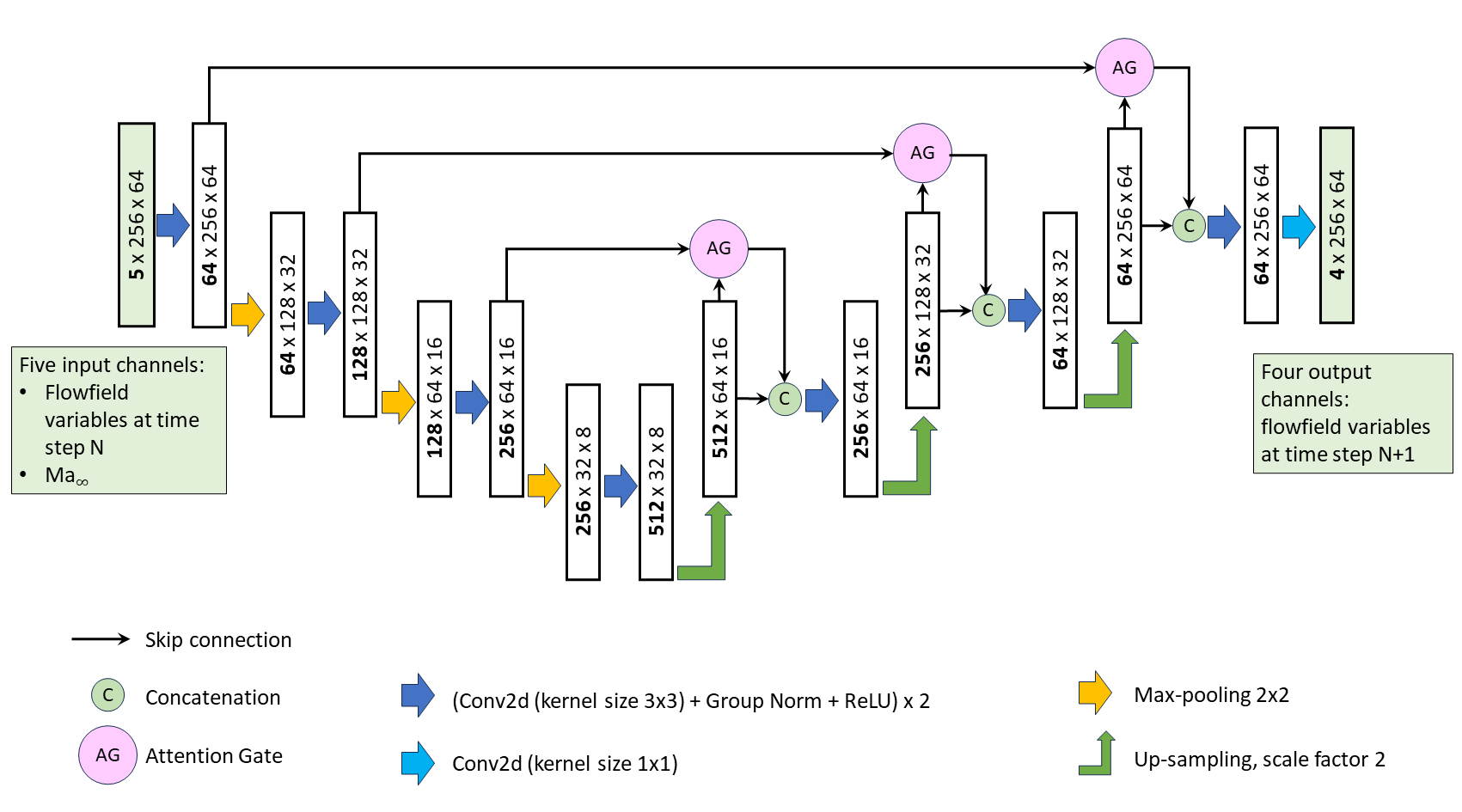}
\caption{The attention U-Net.}
\end{subfigure}

\begin{subfigure}{1.\textwidth}
\centering
\includegraphics[width=\linewidth]{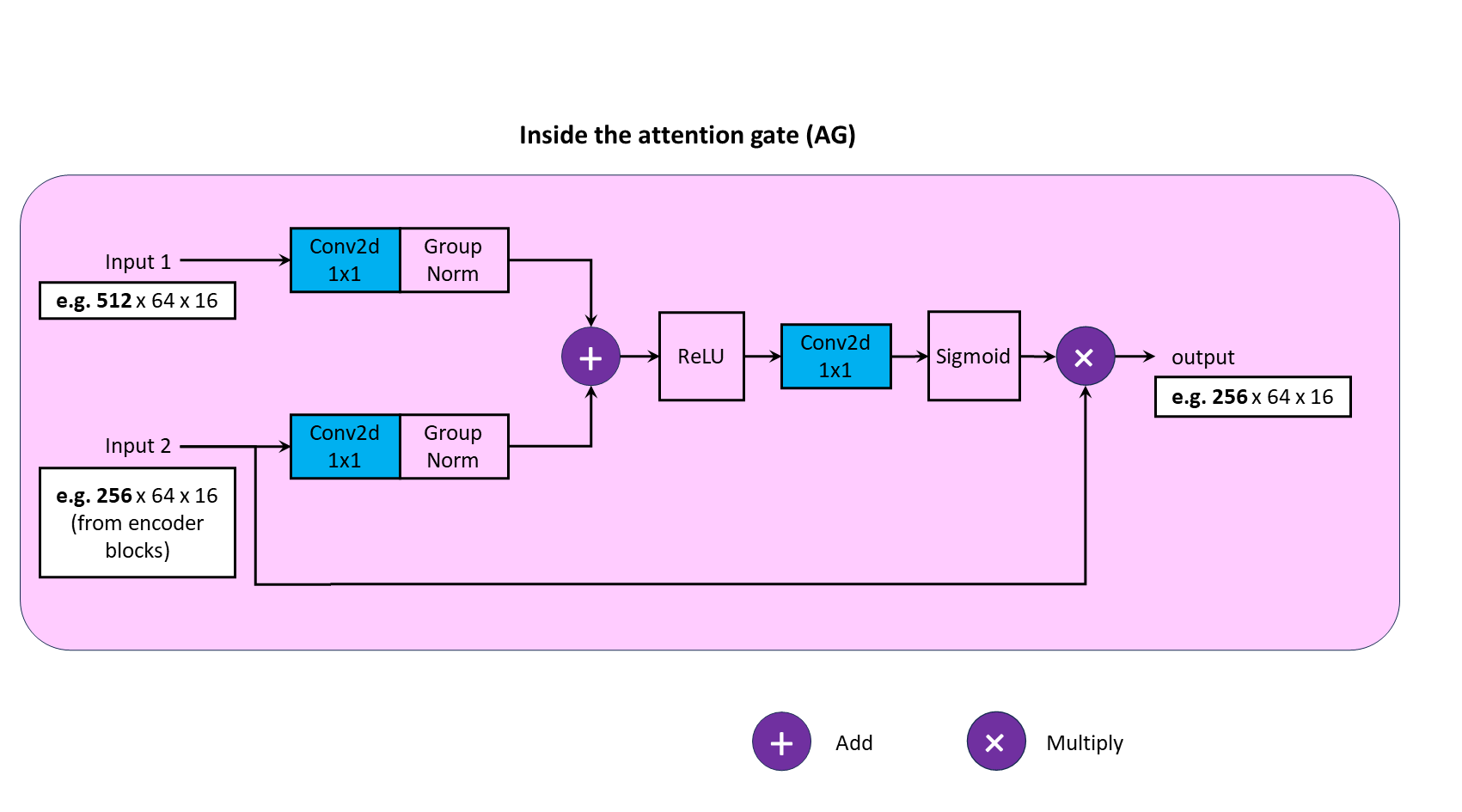}
\caption{The basic schematic of attention gate.}
\end{subfigure}
    \caption{An overview of the attention U-Net architecture with input and output specifications. It contains 8.14 million weights.}
    \label{fig-App:details_about_attentionUnet}
\end{figure}

Figure \ref{fig-App:details_about_attentionUnet} shows an overview of the attention U-Net architecture. It comprises encoder and decoder pathways: the encoder progressively extracts hierarchical features from input images, while attention gates within the architecture selectively emphasize relevant features and suppress irrelevant ones. The decoder pathway then upsamples the features to recover spatial information, with attention mechanisms refining the reconstructed features. During the forward pass, the input flowfield undergoes convolutional operations and attention-based refinement to produce outputs, allowing the model to focus on crucial input regions, thus enhancing its performance in the prediction.
To explore the influence of neural network size on accuracy, multiple training runs are conducted using three distinct network sizes (see table \ref{tab:loss-accumulated}). This is accomplished by adjusting the number of features in each layer.

\begin{figure}
    \centering
    \includegraphics[width=.75\textwidth]{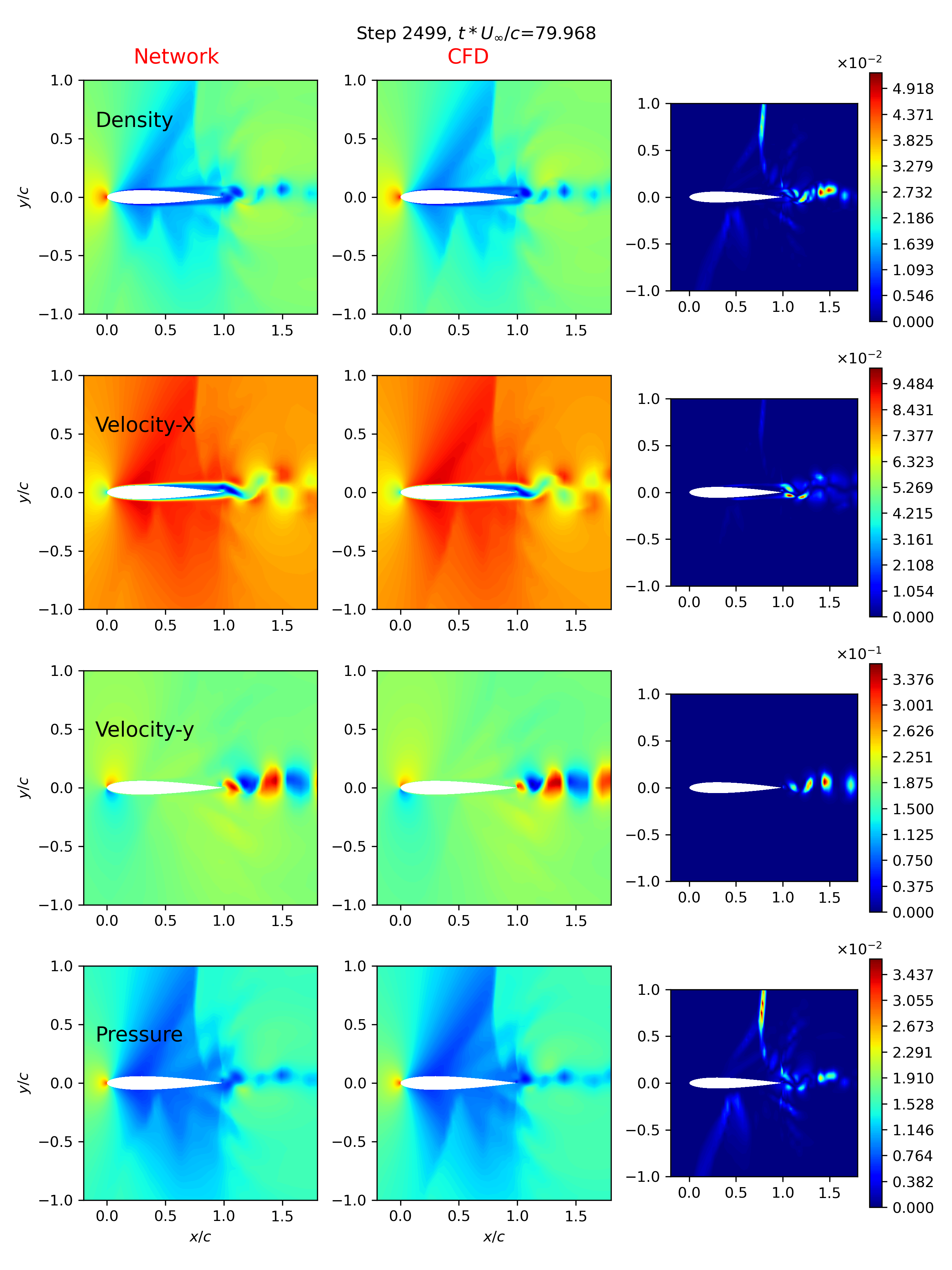}
    \caption{The flowfield from the inference at $M_{\infty}=0.84$ after 2499 prediction steps.}
    \label{fig-App:visualization_DNN_CFD_2499_m0p84}
\end{figure}

Figure \ref{fig-App:visualization_DNN_CFD_2499_m0p84}
illustrates the flow field at $M_{\infty}=0.84$ after 2499 prediction steps, compared with CFD results. The velocity and pressure contour plots reveal minor errors near shock waves and areas of strong compression, as well as in the wake. However, the overall prediction of the flow field structure is remarkably accurate, with the flow field remaining remarkably stable even after 2499 steps.

\bibliographystyle{unsrt}
\bibliography{main}

\end{document}